\documentclass[amsmath,amssymb,aps,prl,twocolumn,showpacs,10pt,floatfix,superscriptaddress]{revtex4-1}
\usepackage{graphicx}
\usepackage{dcolumn}
\usepackage{bm}
\usepackage{dsfont}
\usepackage[utf8]{inputenc}

\begin{document}

\newcommand{\ave}[1]{\left\langle #1\right\rangle}
\newcommand{\wv}[3]{{}_{#1}\ave{#2}_{#3}}
\def\Tr{\mathrm{Tr}}
\def\tr{\mathrm{tr}}
\def\re{\mathrm{Re}}
\def\im{\mathrm{Im}}
\newcommand{\bra}[1]{\left<{#1}\right|}
\newcommand{\ket}[1]{\left|{#1}\right>}
\newcommand{\kett}[1]{|{#1}\rangle}
\newcommand{\braket}[2]{\left<\left.{#1}\right|{#2}\right>}
\newcommand{\expt}[3]{\left<{#1}\left|{#2}\right|{#3}\right>}
\def\pdag{{\phantom{\dagger}}}
\newcommand{\rs}{\rm \scriptscriptstyle}
\def\alphare{\alpha_{\re}}
\def\alphaim{\alpha_{\im}}
\def\alpharez{\alpha_0^{\text{re}}}
\def\alphaimz{\alpha_0^{\text{im}}}
\def\betare{\beta_{\re}}
\def\betaim{\beta_{\im}}
\newcommand{\sgn}[1]{\text{sgn}\left[{#1}\right]}

\title{Dissipation-induced anomalous multicritical phenomena}
\date{\today}

\author{M. Soriente}
\affiliation{Institute for Theoretical Physics, ETH Zurich, 8093 Z{\"u}rich, Switzerland}
\author{T. Donner}%
\affiliation{Institute for Quantum Electronics, ETH Zurich, 8093 Z{\"u}rich, Switzerland}
\author{R. Chitra}
\affiliation{Institute for Theoretical Physics, ETH Zurich, 8093 Z{\"u}rich, Switzerland}
\author{O. Zilberberg}%
\affiliation{Institute for Theoretical Physics, ETH Zurich, 8093 Z{\"u}rich, Switzerland}

\begin{abstract}
We explore the influence of dissipation on a paradigmatic driven-dissipative model where a collection of two level atoms interact with both quadratures of a quantum cavity mode. The closed system exhibits multiple phase transitions involving discrete and continuous symmetries breaking and all phases culminate in a multicritical point. In the open system, we show that infinitesimal dissipation erases the phase with broken continuous symmetry and radically alters the model's phase diagram. The multicritical point now becomes brittle and splits into two tricritical points where first- and second-order symmetry-breaking transitions meet. A quantum fluctuations analysis shows that, surprisingly, the tricritical points exhibit anomalous finite fluctuations, as opposed to standard tricritical points arising in $^3He-\text{}^4He$ mixtures. Our work has direct implications for a variety of fields, including cold atoms and ions in optical cavities, circuit-quantum electrodynamics as well as optomechanical systems.
\end{abstract}

\maketitle
Dissipation can fundamentally influence quantum many-body systems and their phase transitions in often counter-intuitive ways. Prime examples of open quantum many-body systems are interacting light-matter systems where state of the art experiments are able to engineer dissipation channels~\cite{Syassen1329, Baumann2010, Barreiro2011, PhysRevLett.110.035302}. They combine in a unique manner the many-body physics of condensed matter systems with quantum optical tools, including driving and well-controlled dissipation~\cite{Carusotto2013,  Hartmann2008}. In addition to fostering deeper understanding of cooperative phenomena, these systems have potential applications in the realms of quantum computation~\cite{Zoller2001}, laser and maser technologies~\cite{Bohnet2012}, and can potentially generate new states of matter such as light-induced superconductivity~\cite{cavalleri}. New universality classes emerge in driven-dissipative systems~\cite{Tomadin2011, Diehl2010, Nagy2011}, and dissipation can generate topological effects~\cite{Diehl2011}. Concurrently, due to the inapplicability of the framework of equilibrium statistical physics, our understanding of driven-dissipative models remains limited, motivating further studies in this field.
 \begin{figure}[ht!]
    \includegraphics[width=\columnwidth]{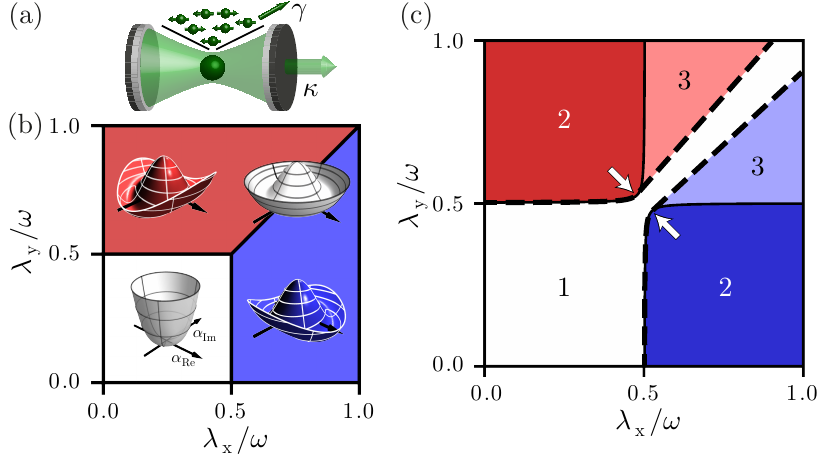}
    \caption{ 
    (a) A schematic illustration of the interpolating Dicke-Tavis-Cummings model [Eq.~\eqref{eq:IDTC}] using a collection of two-level atoms that are strongly coupled to an optical cavity mode with dissipation rate $\kappa$ and atomic decay rate $\gamma$~\cite{Barrett2017}. (b) Phase diagram of the non-dissipative  model [Eq.~\eqref{eq:IDTC}] with $\omega=\omega_c=\omega_a$. The system displays a $\mathds{Z}_2 \times \mathds{Z}_2$-symmetry except for the $\lambda_x=\lambda_y$ Tavis-Cummings line, where a continuous $U(1)$-symmetry emerges~\cite{Baksic2014}. Four distinct phases are possible indicated by their respective free-energy landscapes as a function of the real and imaginary parts of the cavity field $\alphare$ and $\alphaim$, respectively. Quantum phase transitions occur between a normal phase (white) and superradiant phases (blue and red), including a transition between the superradiant phases through the higher-symmetry diagonal Tavis-Cummings line. The phases meet at a multi-critical point. (c) Steady-state phase diagram of the dissipative model [Eq.~\eqref{liouvillian}] for $\gamma = 0$ and $\kappa/\omega= 0.1$. The $U(1)$ high symmetry line and the multi-critical point are washed out by dissipation, replaced by a normal phase sliver that separates the two superradiant phases. Additionally, new regions of coexisting solutions appear (light red and light blue regions). Each region is marked by the number of stable physical solutions~\cite{supmat}. The white arrows point to the new tricritical points where a second-order phase transition line meets a first-order phase transition line. Solid (dashed) lines mark the stability-boundary of the normal (superradiant) phase. }
                \label{fig1}
\end{figure}

Paradigmatic models of driven-dissipative light-matter systems involve multiple spin-like degrees of freedom that are driven and strongly coupled to bosonic cavity modes (see Fig. \ref{fig1}(a))~\cite{Dimer, Domokos, MacDonald}. Such models commonly exhibit quantum phase transitions (QPT) from a normal (NP) to superradiant phases (SP) depending on the coupling between the spins and cavities. Controlled realizations of such models can be achieved in cold atomic quantum gases in high finesse optical cavities~\cite{Baumann2010}. These engineered systems can be used to study quantum phase transitions both in and out of equilibrium in an extremely controlled manner, e.g.,  (i) a $\mathds{Z}_2$ QPT in the so-called driven-dissipative Dicke-model with a single cavity~\cite{Baumann2010, Klinder2015}, (ii) $U(1)$ supersolid symmetry breaking when coupling two cavity modes to the atoms~\cite{Leonard2017}, and (iii) coupling of high-spin atoms to a cavity mode~\cite{Barrett2017}. The latter two cases manifest rich phase diagrams where multiple broken symmetry phases meet at a multicritical point. Similar features are also shared by other fundamental models such as the Lipkin-Meshkov-Glick (LMG) model~\cite{Morrison2008}, or models with QPTs of tunable symmetries~\cite{Baksic2014, Fan2014}. Interestingly, already for the Dicke model, the inevitable coupling to dissipation channels 
was shown to alter the closed system physics~\cite{Keeling2017, Larson2017}. In particular, cavity dissipation was shown to lead to shifts of the critical points and to modifications of critical exponents~\cite{Nagy2015, Kulkarni2013, Brennecke2013}.


Here, we study the connection between the closed and open phase diagram of a paradigmatic driven-dissipative model~\cite{Baksic2014}, hosting a multicritical point and phase transitions breaking discrete and continuous symmetries. Infinitesimal dissipation  dramatically impacts the  model's phase diagram, resulting in rich phenomena, including a splitting of a multicritical point into two tricritical points, coexistence of phases~\cite{Keeling2010}, and relics of the continuous symmetry in rotated order parameters.  In particular,  we analyze the model's quantum fluctuations  and show that these tricritical points  exhibit anomalous finite fluctuations, as opposed to standard tricritical points ~\cite{griffiths}. 

We consider a bosonic cavity mode coupled to $N$ two-level systems described by the Hamiltonian~\cite{Baksic2014, Fan2014}
\begin{align}
\label{eq:IDTC}
H  &= \hbar\omega_c a^\dagger a + \hbar\omega_a  S_z +\\
&+ {\frac{2\hbar\lambda_x}{\sqrt{N}}} S_x (a + a^\dagger)+ {\frac{2\hbar\lambda_y}{\sqrt{N}}} i S_y (a - a^\dagger)\,,\nonumber
\end{align}
where $a^\dagger$ and $a$ are the bosonic creation and annihilation operators of the cavity field, respectively, and the cavity's resonance frequency is $\omega_c$. The collective spin operators $S_\alpha = \sum_{j=1}^N \sigma_\alpha^j$  with $\alpha=x,y,z$ are constituted from the individual Pauli spin operators $\sigma_{\alpha}^{j}$  describing the identical two-level systems with level spacing $\hbar\omega_a$. The two quadratures of the cavity field couple to different projections of the collective spin operators with arbitrary real couplings $\lambda_x$ and $\lambda_y$. Hence, this model interpolates between two ubiquitous light-matter models, the Dicke model~\cite{Dicke,Emary2003} limit when either $\lambda_x=0$ or $\lambda_y=0$ and the Tavis-Cummings model~\cite{TavisCummings} for $\lambda_x=\lambda_y$, and will be dubbed here as the interpolating Dicke-Tavis-Cummings (IDTC) model. A schematic illustration of the model is depicted in Fig.~\ref{fig1}(a). 
\begin{figure}[ht!]
    \includegraphics[width=\columnwidth]{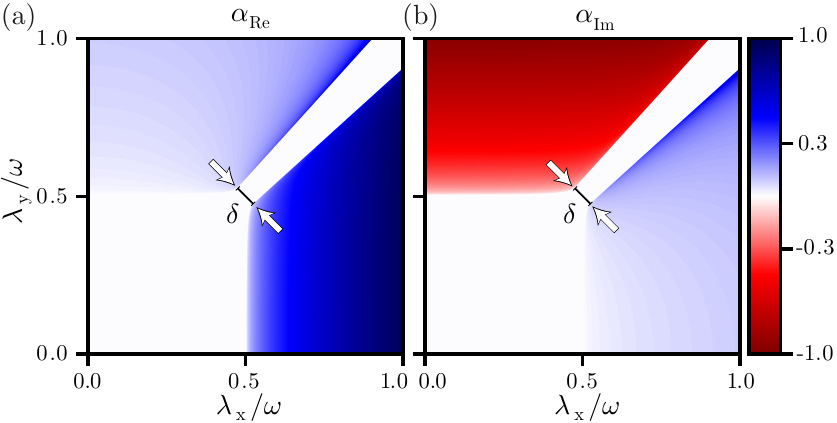}
   	\caption{
    (a) Real and (b) imaginary part of the order
     parameter describing the cavity occupation number,
     $\alphare$ and $\alphaim$, for one of the four 
    possible mean field solutions~\cite{supmat}. Dissipation leads to
	 leakage across the NP diagonal sliver, i.e., the cavity field $\alpha$ becomes 
	complex within the SP regions. Additionally, $\alphaim$ changes sign as a function of 
	$\lambda_x$ and $\lambda_y$, thus, showing that the mean field steady state solutions are rotating 
	within the complex plane. $\delta$ indicates the separation between the two out-of-equilibrium tricritical points~\cite{supmat}. In both plots $\omega=\omega_c=\omega_a$, $\kappa/\omega =0.1$ and 
    $\gamma=0$.}
    \label{fig2}
\end{figure}

The IDTC model has a $\mathds{Z}_2\times \mathds{Z}_2$ symmetry, except for along the diagonal $\lambda_x=\lambda_y$ where it has an enlarged $U(1)$ symmetry, as schematically shown in Fig.~\ref{fig1}(b)~\cite{Baksic2014}. For $\lambda_x,\lambda_y\leq \lambda_{c}\equiv\sqrt{\omega_c\omega_a}/2$, the system has a trivial ground state, dubbed normal phase (NP), that is comprised of an empty cavity and all two-level system in their respective ground states, oriented along the $z$-axis. Fixing one of the couplings below criticality $\lambda_i\leq \lambda_c$ and taking the other above it $\lambda_j> \lambda_c$ with $i,j\in\{x,y\}$ and $i\neq j$, the system undergoes a $\mathds{Z}_2$-breaking Dicke-like transition from the NP to a superradiant phase (SP), where the cavity features a finite mean population, $\ave{a}\neq 0$, and the two-level systems are on average oriented away from the $z$-axis. Beyond the SP threshold and along the diagonal, $\lambda_x=\lambda_y$, the $U(1)$ symmetry is spontaneously broken marking the Tavis-Cummings QPT~\cite{Baksic2014}. The hallmark of the IDTC model is the appearance of a multicritical point  at $\lambda_x=\lambda_y = \lambda_c$, where all phases meet and the symmetry of the Hamiltonian changes from a discrete to a continuous symmetry~\cite{Baksic2014}.

The relevant complex order parameter which captures these transitions is $\langle a \rangle = \sqrt{N} \alpha$, where $\alpha= \alphare+ i \alphaim$. At a Dicke-like phase transition either $\alphare \neq 0$ and $\alphaim = 0$, or $\alphare= 0$ and $\alphaim \neq 0$. Along the Tavis-Cummings line, the $U(1)$-symmetry is broken and both $\alphare, \alphaim \neq 0$. The Ginzburg-Landau  energy potential of the order parameter is schematically plotted in Fig.~\ref{fig1}(b) and was calculated in Ref.~\cite{Baksic2014}. It shows (i) a single minimum in the normal phase, (ii) two minima along either  the real or the imaginary axis marking the $\mathds{Z}_2$ Dicke-like symmetry breaking, and (iii) a ``sombrero-hat'' potential with an enlarged $U(1)$ symmetry on the diagonal.

The closed system phase diagram [Fig.~\ref{fig1}(b)], however, fundamentally changes if one includes dissipation in the model, which will be relevant for any experimental implementation of such a system. In the presence of both cavity and global atomic dissipation, the driven and dissipative nature of the system is described by a Liouvillian equation for the density matrix $\rho_{\rm sys}$ of the system~\cite{vogel2006}
\begin{eqnarray}
\frac{d \rho_{\rm sys}}{dt} &=& - \frac{i}{\hbar} [H(t), \rho_{\rm sys}]  + \kappa[ 2 a \rho_{\rm sys} a^\dagger - \{ a^\dagger a , \rho_{\rm sys}\}]  \nonumber \\ &&+ 
\frac{\gamma}{N} [ 2 S_- \rho_{\rm sys} S_+ -\{S_+ S_-, \rho_{\rm sys}\}]\,,
\label{liouvillian}
\end{eqnarray}
where $S_{\pm}=S_{x}\pm iS_{y}$ are ladder operators. The first term on the r.h.s. describes the standard Hamiltonian evolution while the other two terms represent the Markovian dissipation for both the cavity and the collective spin in Lindblad form with rates $\kappa$ and $\gamma$, respectively. It should be noted that the master equation~\eqref{liouvillian} is valid in the rotating frame of driven systems with weak cavity-spin coupling, cf.~Refs.~\cite{Baumann2010,Barrett2017,Leonard2017}. The ultrastrong coupling regime  should  generally be investigated using  dressed operators~\cite{Miyashita2014,Lolli2015}. Here, we use Eq.~\eqref{liouvillian} and discuss the validity of our results in the strong coupling regime in~\cite{dressed}. In the following, we set $\gamma=0$ and focus mainly on cavity dissipation. 
 
The mean-field equations governing the different ordered phases can be derived from Eq.~\eqref{liouvillian},
\begin{align}
\omega_c\alphaim - \kappa \alphare -2\lambda_y Y &= 0\,, \label{MF1}\\
\omega_c \alphare +\kappa\alphaim +  2 \lambda_x X & = 0\,,\label{MF2}\\
 \omega_a Y + 4 \lambda_y \alphaim  Z  &= 0\,, \label{MF3}\\
 \omega_a X - 4 \lambda_x \alphare Z  &= 0\,,\label{MF4}
\end{align}
where we defined $\langle S_x\rangle = NX$, $\langle S_y\rangle = NY$ and $\langle S_z\rangle = NZ$ and have taken the steady-state limit. Additionally, for the case of the global dissipation considered here, we have the spin-conservation law $X^2 + Y^2 + Z^2 = 1/4$ that is used to solve the mean-field equations analytically~\cite{supmat}.

Solving Eqs.~\eqref{MF1}-\eqref{MF4}, we find that dissipation stabilizes a richer phenomenology inaccessible in the closed system paradigm, with multiple bifurcations  and coexisting many-body phases, see Fig.~\ref{fig1}(c)~\cite{supmat}. Specifically, the mean-field equations can be manipulated to obtain an equation for $Z$: $\kappa^2 \omega_a^2+ \left[\left(8 Z \lambda_x^2 + \omega_c \omega_a\right) \left(8 Z \lambda_y^2 + \omega_c \omega_a\right)\right] =0$~\cite{supmat}. All valid solutions are then subject to a constraint $4 \kappa^2 \lambda_x^2 \lambda_y^2 \le (\lambda_x^2 -\lambda_y^2)^2\omega_c^2$~\cite{supmat}. However, only a subset of these solutions are compatible with the aforementioned spin-conservation law. Note that, the spin conservation is trivially violated in the presence of single spin dephasing leading to the destruction of Dicke superradiant phases,~\cite{Keeling2017,Torre2016}, which is beyond the scope of this work.

The implications of this constraint are numerous, for example, (i) it is violated at the Tavis-Cummings line, $\lambda_x=\lambda_y$, implying that the $U(1)$ QPT is destroyed by dissipation, as can be easily seen using adiabatic elimination~\cite{supmat} [cf.~Ref.~\cite{Larson2017}]; (ii) it is additionally violated in a $\kappa$-dependent sliver around the Tavis-Cummings line, see Fig.~\ref{fig1}(c). Consequently, the NP percolates through this sliver beyond the multicritical region of the closed system. Away from the sliver, solving the mean-field equations leads to multiple stable and unstable solutions~\cite{supmat}. In particular, the independent Dicke-like phases boast two stable solutions corresponding to the $\mathds{Z}_2$-broken SP states, as well as an unstable NP solution, a feature  seen  also in ~\cite{Keeling2017}. Interestingly, in the regime where both couplings are above $\lambda_c$, coexistence of both stable SP and NP states appears, supported by two unstable solutions, see Fig.~\ref{fig1}(c) and~\cite{supmat}. 

Dissipation leads to additional important features in the IDTC as evinced by Fig.~\ref{fig2} where the order parameters, $\alphare$ and $\alphaim$, for one of the stable  symmetry broken solutions are  plotted. Clearly,  $\alphare$ and $\alphaim$ leak across the NP diagonal sliver, unlike the nondissipative case [cf.~Fig.~\ref{fig1}]. Consequently, the cavity field $\alpha$ is complex within the $\mathds{Z}_2$ SP regions, which can be attributed to  the remnant memory of the underlying $U(1)$ symmetry in the problem. Similarly, $\alphaim$ changes sign as a function of $\lambda_x$ and $\lambda_y$, thus showing that the mean field solutions are rotating within the complex plane. Importantly, the order parameter components, $\alphare$ and $\alphaim$ evolve continuously from zero across the Dicke-like phase boundaries, but show a discontinuous behavior along the dissipation-induced NP sliver. Therefore, along the sliver edges, there are two second-order transition lines, which morph into two first-order transition lines in the vicinity of the nondissipative multicritical point, $\lambda_x=\lambda_y=\lambda_c$, i.e., these lines meet at new dissipation-induced tricritical points,  which separate the continuous and discontinuous symmetry breaking transitions in the system, see Figs.~\ref{fig1}(c) and \ref{fig2}. For $\omega_a=\omega_c=\omega$, we find that the separation between the two out-of-equilibrium tricritical points scales linearly with $\kappa$ for $\kappa<<\omega$ and is given by $\delta = \omega \sqrt{1 + ( \kappa/\omega) ^2 -\sqrt{1 + ( \kappa/\omega) ^2} } \approx \kappa/\sqrt{2}$~\cite{supmat}.

\begin{figure*}[ht!]
    \includegraphics[width=\textwidth]{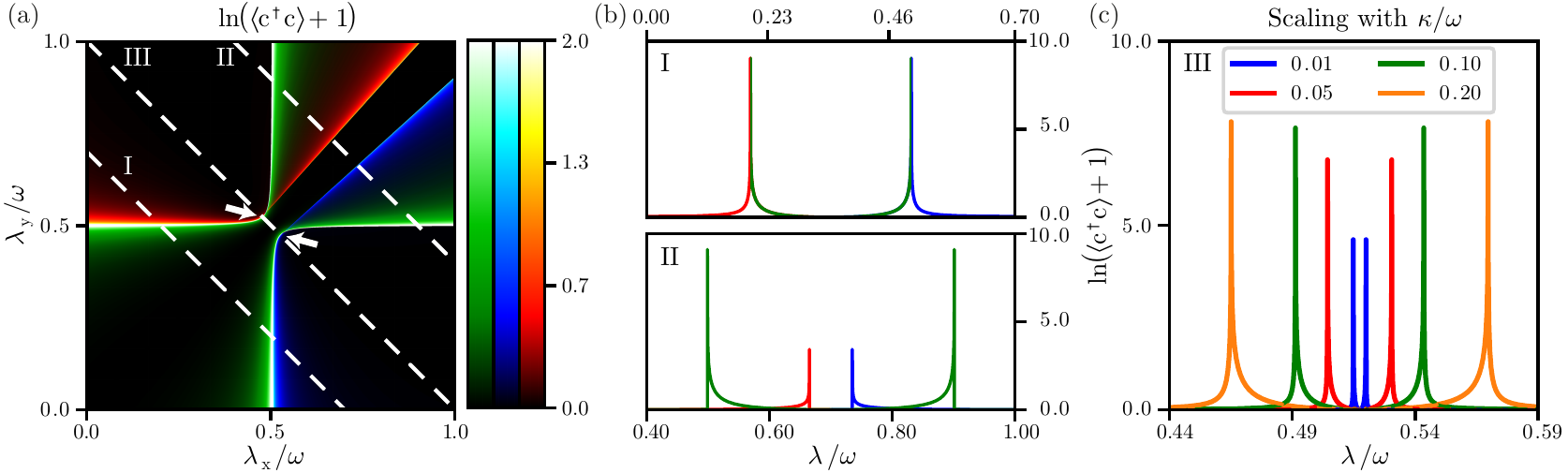}
    \caption{
    (a) Density plot of the photon number fluctuations calculated on top of the stable mean-fields solutions. $\ln\left\{\langle c^\dagger  c\rangle +1 \right\}$ is plotted for clarity. Green, blue and red refer to fluctuations on top of the normal phase and the two superradiant phases, respectively. The standard Dicke-like transitions exhibit expected continuous second-order transition lines, which morph into two first-order lines marking the regions of coexisting solutions. There are no fluctuations along the Tavis-Cummings line [cf.~Ref.~\cite{Larson2017}], and due to dissipation, the zero-fluctuation region broadens in the IDTC phase space. (b) Cuts with couplings $\lambda$ along lines $\rm I$ (top panel) and $\rm {II}$ (bottom panel) of plot (a). Top panel: Dicke-like transitions where the fluctuations continuously diverge on both sides of the critical point with exponent $1$. Bottom panel: one sided first-order phase transitions marking the boundaries of the NP (green) and the two SPs (blue and red). (c) Cut with couplings $\lambda$ along the line $\rm {III}$ of plot (a) for different values of $\kappa/\omega$. As $\kappa$ decreases, the fluctuations decreases and the separation between the two SPs shrinks, eventually recovering the closed system behavior and the multicritical point. In all plots $\omega=\omega_c=\omega_a$ and $\gamma = 0$. In (a) and (b) $\kappa/\omega = 0.1$.}
    \label{fig3}
\end{figure*}

To summarize, the dissipation renders the multicriticality of the IDTC model brittle. Nevertheless, signatures of the Tavis-Cummings $U(1)$ symmetry manifest in the splitting of the multicritical point into two new tricritical points with coexisting phases, and the two closed-system SP phases are separated by an emerging NP sliver. Importantly, the multicriticality is sensitive to even infinitesimally small cavity dissipation. This is radically different from standard driven-dissipative systems such as the Dicke model, where cavity dissipation or a global atomic dissipation engender a small modification of phase boundaries~\cite{Domokos, Dimer}.  

A more thorough characterization of the phase diagram is provided by an analysis of the steady state fluctuations and their scaling beyond mean field.  In all regimes of the parameter space, the fluctuation Hamiltonian reduces to a problem of two coupled linear oscillators~\cite{supmat}. In the thermodynamic limit (large-$N$), it takes the generic form~\cite{Baksic2014}
\begin{align}
\label{fluc}
H_{\rm fl} =& \hbar\omega_c c^\dagger c +\hbar\Omega_a d^\dagger d \\
&+\left(\Gamma_1 c^\dagger d + \Gamma_2 c d +\Gamma_3  {d^\dagger}^2 + h.c.\right)\nonumber\,,
\end{align}
where we have defined $a=\alpha\sqrt{N}+c$ with $c$ the bosonic cavity fluctuation operator, and have used the Holstein-Primakoff representation for the spins, $S_+=  b^\dagger \sqrt{N - b^\dagger b}$ and $S_z= -\frac{N}{2} + b^\dagger b$. Here, $b=\beta\sqrt{N}+d$ is a bosonic annihilation operator describing deviations away from the $z$-axis, with $\beta=(X-iY)/\sqrt{1/2 - Z}$ its mean value and $d$ the atomic fluctuation operator. The parameters $\Omega_a, \Gamma_1, \Gamma_2$ and $\Gamma_3$ are determined by the mean-field solutions~\cite{Baksic2014,supmat}.

A normal mode transformation on Eq.~\eqref{fluc} yields the excitation spectra of the problem as was studied for the closed system case in Ref.~\cite{Baksic2014}. To analyze the nature of the phase transitions in the open system dynamics, we calculate, using the Liouvillian \eqref{liouvillian}, the time evolution of the equal-time two-operator correlation functions. The resulting equations of motion form a closed set of ten coupled differential equations for the correlators, $ \langle c^\dag c\rangle$,  $ \langle c^\dag c^\dag\rangle$,  $ \langle c c\rangle$,  $\langle d^\dag d\rangle$, $\langle d^\dag d^\dag\rangle$, $\langle dd\rangle$, $\langle c^\dag d\rangle$,  $\langle c^\dag d^\dag\rangle$,  $\langle c d\rangle$, and $\langle d^\dag c\rangle$~\cite{supmat, Oztop2012}. In the steady state, this set conforms to a set of linear inhomogeneous equations, which can then be solved for $\lambda_x \ne \lambda_y$ to obtain the various correlators~\cite{supmat}. All correlators display similar features and in the following we shall focus on the photon number correlator $\langle c^\dagger  c\rangle $, see Fig.~\ref{fig3}.

It is instructive to study the expectation value for the photon number in the normal phase. For $\omega_a=\omega_c=\omega$, it is given by 
 \begin{align}
\label{photon}
\langle c^\dagger  c\rangle=& \frac{(\lambda_x^2 - \lambda_y^2)^2}{2[\kappa^2 \lambda_x\lambda_y + (\lambda_x+\lambda_y)^2\omega^2]} \\
&\times\frac{\omega^2 ( \omega^2 + \kappa^2 + 4 \lambda_x \lambda_y)}{[16 \lambda_x^2\lambda_y^2 +\omega^2(\omega^2 +\kappa^2 - 4(\lambda_x^2+\lambda_y^2))]}\,.\nonumber
\end{align}
In the Dicke-limit, where $\lambda_i \to \lambda_c$ while $\lambda_j=0$ with $i\neq j$, the photon number diverges with an exponent of $1$, as expected~\cite{Nagy2011}. Note, however, that away from the Dicke-limit, the critical regime of the Dicke-like transitions shrinks. This is consistent with the fact that the multicritical point is expected to have zero fluctuations~\cite{supmat}. For the case $\lambda_x=\lambda_y$, we find that the set of ten equations is no longer invertible. However, a reduced solvable set exists for which the photon fluctuations vanish. This is a remarkable result which shows that even weak cavity dissipation which technically preserves the length of the spin destroys an ordered phase.

The specific features of the QPTs in the IDTC can be inferred by considering photon number fluctuations along three representative cuts in the $\lambda_x- \lambda_y$ plane shown  in Fig.~\ref{fig3}(a). Along $\rm I$, where at least one of the couplings is below $\lambda_c$ (top panel of Fig.~\ref{fig3}(b)), as expected for a continuous  Dicke-like QPT, cavity fluctuations diverge continuously across both sides of the transition with critical exponent $1$. Across cut $\rm {II}$ (bottom panel of Fig.~\ref{fig3}(b)),  where  either $\lambda_x$ or $\lambda_y$ is greater than $\lambda_c$, the loss of the broken $U(1)$ phase results in discontinuous first-order transitions between the NP and the two SP phases concomitant with coexistence regions. Cavity fluctuations, though enhanced, remain finite and exhibit a jump across the phase boundaries. The widths of the coexistence regimes effectively indicate the size of hysteresis loops that will appear under scans of the couplings. In contrast to standard tricritical points that arise in systems as diverse as $^3He-\text{}^4He$ mixtures~\cite{griffiths} and high-Tc superconducting vortex lattices~\cite{vortex}, the fluctuations at the two out-of-equilibrium tricritical points of the IDTC remain finite. The scaling of these fluctuations with dissipation is illustrated in Fig.~\ref{fig3}(c) for cut $\rm {III}$, which shows that fluctuations diminish as the two tricritical points approach the original multicritical point.

We have shown that weak dissipation can dramatically alter the paradigm of standard continuous symmetry breaking phase transitions in a model system exhibiting multicriticality. Additionally, the dissipation induced tricritical points are characterized by anomalous quantum fluctuations. We expect our results to be qualitatively valid for other dissipation channels provided the spin is conserved. Extending our work to variants of the IDTC with (i) higher-spin systems, (ii) negatively detuned frequencies where interesting oscillatory behavior are expected [cf.~Ref.~\cite{Barrett2017}], (iii) additional cavity fields [cf.~Ref.~\cite{Keeling2018}], (iv) dephasing and non spin-conserving dissipation, and (v) quenched dynamics will further reinforce our predictions for existing experiments~\cite{Leonard2017,Barrett2017}. Our work also motivates a study of potential brittle multicritical phenomena in quantum engineered systems, and in out-of-equilibrium matter systems as well as the influence of non-Markovian noise on such phase diagrams.



We would like to thank I. Carusotto,  A. Imamoglu, J. Keeling, and M. Landini for useful discussions. We acknowledge financial support from the Swiss National Science Foundation (SNSF), Division 2 and through the SNSF DACH-project ``Quantum Crystals of Matter and Light''.

\newpage
\cleardoublepage
\setcounter{figure}{0}
\renewcommand{\figurename}{Supplementary Material Figure}

\onecolumngrid
\begin{center}
\textbf{\normalsize Supplemental Material for:}\\
\vspace{3mm}
\textbf{\large Dissipation-induced anomalous multicritical phenomena}
\vspace{4mm}

{ Matteo\ Soriente,$^{1}$ Tobias Donner,$^{2}$ R. Chitra,$^{1}$ and Oded Zilberberg$^{1}$}\\
\vspace{1mm}
\textit{\small $^{1}$Institute for Theoretical Physics, ETH Z\"urich, 8093 Z\"urich, Switzerland\\
$^{2}$Institute for Quantum Electronics, ETH Zurich, 8093 Z{\"u}rich, Switzerland}

\vspace{5mm}
\end{center}
\setcounter{equation}{0}
\setcounter{figure}{0}
\setcounter{table}{0}
\setcounter{page}{1}
\makeatletter
\renewcommand{\bibnumfmt}[1]{[S#1]}
\renewcommand{\citenumfont}[1]{S#1}

\setcounter{enumi}{1}
\renewcommand{\theequation}{\Roman{enumi}.\arabic{equation}}

\twocolumngrid

\section{I. Adiabatic elimination}  
For the case of symmetric coupling ($\lambda = \lambda_x = \lambda_y$), the Hamiltonian can be written as
\begin{align}
H =& -\hbar \omega_c a^\dagger a + \hbar \omega_a S_z + \lambda (a S_+ + a^\dagger S_-)\,.
\end{align}
From this we derive the equations of motion for the expectation values of the operators $S_z$ and $a$ including cavity decay at rate $\kappa$,
\begin{align}
 \langle\dot{S}_z\rangle=& 2i \lambda \left( \langle a^\dagger\rangle \langle S_-\rangle - \langle a\rangle  \langle S_+\rangle \right)\,,\\
 \langle\dot{ a}\rangle =& i \omega_c \langle a\rangle - i \lambda \langle S_-\rangle - \kappa \langle a\rangle\,,
\end{align}
where the factorization $\langle a S_i\rangle = \langle a\rangle\langle S_i\rangle$ has been imposed.
Consdering the steady state for the cavity field ($\dot{\langle a\rangle}=0$) gives
\begin{align}
  \label{a_adiabatic} \langle a\rangle =& -\frac{2\lambda \langle S_-\rangle}{\omega_c - i \kappa}\\
  \label{adag_adiabatic}   \langle a^\dagger\rangle =& -\frac{2\lambda \langle S_+\rangle}{\omega_c + i \kappa}\,.
\end{align}

We then look at the steady state for the spin population, i.e. $\langle\dot{S}_z\rangle=0$, which results in
\begin{align}
  \langle a\rangle \langle S_+\rangle = \langle a^\dagger\rangle \langle S_-\rangle\,,
\end{align}
 i.e., in detailed balance for the excitations in the system.
Making use of the expressions (\ref{a_adiabatic}) and (\ref{adag_adiabatic}) for the light fields, we find
\begin{align}
 \label{detailed_balance} \frac{\langle S_-\rangle \langle S_+\rangle }{\omega_c-i\kappa} =   \frac{\langle S_-\rangle \langle S_+\rangle}{\omega_c + i\kappa}\,.
\end{align}
Any finite cavity decay $\kappa \neq 0$ thus gives rise to the breakdown of detailed balance, leaving only the trivial solution $\langle a\rangle=0$, and therefore hindering the Tavis-Cummings phase transition.

\section{II. Mean-field solution} 
\setcounter{enumi}{2}
\setcounter{equation}{0}
In order to solve the mean-field Eqs.~(3)-(6) in the main text, we first obtain expressions for $X$ and $Y$ as functions of $\alphare$, $\alphaim$ and $Z$, for the case of vanishing atomic dissipation ($\gamma=0$). Eqs.~(3) and (4) lead to
\begin{align}
\label{x1} X=& -\frac{\omega_c\alphare + \kappa\alphaim }{2 \lambda_x}\,,\\
\label{y1} Y=& \frac{\omega_c\alphaim - \kappa\alphare }{2 \lambda_y}\,,
\end{align}
while Eqs.~(5) and (6) to
\begin{align}
\label{x2} X=& \frac{4 \lambda_x\alphare Z }{\omega_a}\,\\
\label{y2} Y=& -\frac{4 \lambda_y\alphaim Z }{\omega_a}\,.
\end{align}
Substituting (\ref{x2},~\ref{y2}) into (\ref{x1},~\ref{y1}), we obtain a homogeneous system for $\alphare, \alphaim$ parametrized by $Z$. Requiring that the system has non-trivial solutions, we find the constraint on $Z$ presented in the main text. This constraint yields two solutions 
\begin{align}
\label{zetapm}
Z_{\pm}=& -\frac{ \left(\lambda_x^2+\lambda_y^2\right)\omega_c \pm \sqrt{\left(\lambda_x^2-\lambda_y^2\right)^2\omega_c^2-4\kappa^2\lambda_x^2\lambda_y^2}}{16\lambda_x^2\lambda_y^2}\omega_a\,.
\end{align}
We shall find it useful in the following to define $Z_{\pm}=\mathcal{A_{\pm}}\omega_a$. 

We insert the solution $Z_{\pm}$, as well as, Eqs.~\eqref{x1}-\eqref{y2} into the spin length normalization constraint, $X^2+Y^2+Z^2=1/4$, and find two inhomogeneous quadratic equations for $\alphare$ and $\alphaim$,
\begin{align}
\label{alre} c=&a_1 \alphare^2 + a_2 \alphaim^2 \,\\
\label{alim} c=&b_1 \alphare^2 + b_2 \alphare \alphaim + b_3 \alphaim^2\,
\end{align}
with coefficients depending on which solution of $Z$-constraint is taken, $a_1=16\mathcal{A}_\pm^2\lambda_x^2$, $a_2=16\mathcal{A}_\pm^2\lambda_y^2$, $b_1=\left(\kappa^2/\lambda_y^2+\omega_c^2/\lambda_x^2\right)/4$, $b_2=\kappa\omega_c\left(1/\lambda_x^2-1/\lambda_y^2\right)/2$, $b_3=\left(\kappa^2/\lambda_x^2+\omega_c^2/\lambda_y^2\right)/4$, and $c=1/4 - \mathcal{A}_\pm^2 \omega_a^2$.

Equations \eqref{alre} and \eqref{alim} are solved at once yielding four possible solutions. We note, however, that our procedure has increased the space of solutions, since we have moved from a first-order system for $\alphare$ and $\alphaim$ to a second-order one. This means that not all of the obtained solutions are \textit{physical}. Imposing that the solutions preserve real spin order parameters and that the total spin is conserved, we find that the validity of the solutions is dependent on the coupling strengths $\lambda_x, \lambda_y$. Specifically, we find that two of the solutions are physical for $\lambda_x > \lambda_y$ while the other two are physical for $\lambda_y > \lambda_x$. This is readily combined and the 
%
solutions can be written as
\begin{align}
&\alphare=\\
&\resizebox{1.\hsize}{!}{$\displaystyle \pm\sqrt{c}\sqrt{\frac{2 a_2^2 b_1 + a_2 b_2^2 - 2 a_2 b_3 (a_1 + b_1) + 2 a_1 b_3^2 + \sgn{\lambda_y - \lambda_x} a_2 |b_2| \sqrt{b_2^2 - 4 (a_1 - b_1)(a_2 - b_3) }}
{2 (a_2^2 b_1^2 + a_1^2 b_3^2 + a_1 a_2 (b_2^2 - 2 b_1 b_3)) }}
\,,$}\nonumber\\
&\alphaim=\frac{b_2^2 - \sgn{\lambda_y - \lambda_x} |b_2|\sqrt{4 (b_1 - a_1)(a_2 - b_3) + b_2^2}}{2 b_2 (a_2 - b_3)}\alphare\,.
\end{align}
Remarkably, in the respective regions of validity, the solutions are the opposite of one another, reflecting the underlying broken $\mathds{Z}_2$ symmetry.

\section{III. Stability analysis}
\setcounter{enumi}{3}
\setcounter{equation}{0} 

In the previous section, we have obtained solutions to the mean-field equations of our model [cf.~Eqs.~(3)-(6) in the main text]. These solutions, however, can be stable or unstable. To check the linear stability of our solutions, we employ standard stability analysis~\cite{Bhaseen2012}. In particular, we find the stability matrix associated with our system of equations and diagonalize it at a given solution. Once we have its eigenvalues, we study their behavior as a function of the coupling strengths, $\lambda_x, \lambda_y$. If at least one eigenvalue has non-vanishing positive real part then the solution is unstable.

More rigorously, we expand the order parameters to linear order in small 
fluctuations, i. e., $\alpha = \alpha_0 + \delta\alpha, X = X_0 + \delta X, Y = Y_0 + \delta Y, Z = Z_0 + \delta Z$. Plugging this ansatz into the mean-field equations, we obtain
\begin{align}
  \dot{\delta\alpha} & = - 2 \kappa \big(\alpha_0 + \delta\alpha \big) - i \omega_0\big(\alpha_0 + \delta\alpha\big) +  \\
  & \quad \,- 2 i \lambda_x \big(X_0 + \delta X\big) - 2 \lambda_y \big(Y_0 + \delta Y\big)\,, \nonumber \\
  \dot{\delta X} & = - \omega_z \delta Y + 2 i \lambda_y \big(\alpha_0 - \alpha_0^*\big) \delta Z +   \\
  & \quad \, + 2 i \lambda_y Z_0 \big(\delta\alpha - \delta\alpha^*\big) \,,\nonumber \\
  \dot{\delta Y} & = \omega_z \delta X - 2 \lambda_x \big(\alpha_0 + \alpha_0^*\big) \delta Z \\
  & \quad \, - 2 \lambda_x Z_0 \big(\delta\alpha + \delta\alpha^*\big)\,, \nonumber \\
  \dot{\delta Z} & = - 2 \lambda_x \big(\alpha_0 + \alpha_0^*\big) \delta Y - 2 \lambda_x Y_0 \big(\delta\alpha + \delta\alpha^*\big) +  \\
  & \quad \, + 2 i \lambda_y \big(\alpha_0 - \alpha_0^*\big)\delta X + 2 i \lambda_y X_0 \big(\delta\alpha - \delta\alpha^*\big) \,,\nonumber
\end{align}
where $\alpha_0, X_0, Y_0, Z_0$ are the mean-field steady-state order parameters. 

We use the steady-state solutions from the previous section and obtain
\begin{align}
  & \dot{\delta\alphare} =  - \kappa \delta\alphare + \omega_0 \delta\alphaim - 2 \lambda_y \delta Y\,, \\
  & \dot{\delta\alphaim} =  - \kappa \delta\alphaim - \omega_0 \delta\alphare - 2 \lambda_x \delta X\,, \\
  & \dot{\delta X} =  - \omega_z \delta Y - 4 \lambda_y \alphaimz \delta Z - 4 \lambda_y Z_0 \delta\alphaim\,, \\
  & \dot{\delta Y} =  \omega_z \delta X - 4 \lambda_x \alpharez \delta Z - 4 \lambda_x Z_0 \delta\alphare\,, \\
  & \dot{\delta Z} =  - 4 \lambda_x \alpharez \delta Y - 4 \lambda_x Y_0 \delta\alphare + \\
  & \qquad\,\, - 4 \lambda_y \alphaimz \delta X - 4 \lambda_y X_0 \delta\alphaim\,. \nonumber
\end{align}
To reduce the number of equations of our system, we linearize the spin constraint, $X^2 + Y^2 + Z^2 = 1/4$, which allows us to get rid of the $\delta Z$ dependence:
\begin{equation}
  \delta Z = - \frac{X_0 \delta X + Y_0 \delta Y}{Z_0}\,.
\end{equation}
Substituting the linearized condition into the previous equations, we can rewrite them in matrix form as follows
\begin{equation}
  \begin{pmatrix} \dot{\delta\alphare} \\ \dot{\delta\alphaim} \\ \dot{\delta X} \\ \dot{\delta Y} \end{pmatrix} = 
  M\begin{pmatrix} \delta\alphare \\ \delta\alphaim \\ \delta X \\ \delta Y \end{pmatrix}\,,
\end{equation}
where the matrix\scriptsize
\begin{equation}
\label{stabilitymatrix}
  M \equiv 
  \begin{pmatrix}   - \kappa    &     \omega_0  &               0                 &           - 2 \lambda_y               \\
            - \omega_0    &     - \kappa  &           - 2 \lambda_x               &               0                 \\
              0     & - 4 \lambda_y Z_0 &     4 \lambda_y \alphaimz \frac{X_0}{Z_0}         & 4 \lambda_y \alphaimz\frac{Y_0}{Z_0} - \omega_z \\
          - 4 \lambda_x Z_0   &       0     & 4 \lambda_x \alpharez\frac{X_0}{Z_0} + \omega_z &       4 \lambda_x \alpharez \frac{Y_0}{Z_0}
  \end{pmatrix}\,,
\end{equation}\normalsize
is the \emph{stability matrix}. We now diagonalize this matrix and check the positivity condition for the real part of the eigenvalues. The $\lambda_x, \lambda_y$ regions for which the real part of an eigenvalue is zero define the stability lines of Fig. 1(c) in the main text.

Performing this analysis, we find that the nontrivial solutions associated with the $Z_-$ constraint are always stable. The ones associated with the $Z_+$ constraint, even though physical, are never stable. Interestingly, performing the stability analysis for the normal phase, we find that it is stable when both couplings are above threshold.


\section{IV. Separation between the two out-of-equilibrium tricritical points}
\setcounter{enumi}{4}
\setcounter{equation}{0} 

We first identify the two tricritical points. This can be done in different ways, e.g. considering the constraints in Eq.~\eqref{zetapm} or the stability conditions, we chose to use the constraint $Z_-$. We can restrict to the normal phase without loss of generality. The tricritical points are then given by the intersection of two curves, namely\small
\begin{align}
\left(\frac{ -\left(\lambda_x^2+\lambda_y^2\right)\omega_c + \sqrt{\left(\lambda_x^2-\lambda_y^2\right)^2\omega_c^2-4\kappa^2\lambda_x^2\lambda_y^2}}{16\lambda_x^2\lambda_y^2}\omega_a\right)^2 & = \frac{1}{4} \nonumber \\
\left(\lambda_x^2-\lambda_y^2\right)^2\omega_c^2-4\kappa^2\lambda_x^2\lambda_y^2 & = 0\,. \nonumber
\end{align}\normalsize
The first one is just the spin constraint in the normal phase, $Z^2=1/4$, while the second one gives the boundary for which the square root becomes imaginary and therefore non physical. We solve this system for $\lambda_x$ and $\lambda_y$ and find that among all the possible solutions the tricritical points are given by
\begin{align}
\label{tricriticalpoints}
\begin{pmatrix} \lambda_{x1} \\ \lambda_{y1} \end{pmatrix} & = \begin{pmatrix} \frac{1}{2} \sqrt{-\frac{\sqrt{\kappa ^2 \omega_a^2 \left(\kappa ^2+\omega_c^2\right)}}{\omega_c}+\frac{\kappa ^2 \omega_a}{\omega_c}+\omega_c \omega_a} \\ \frac{1}{2} \sqrt{\frac{\kappa  \omega_a \left(\sqrt{\kappa ^2+\omega_c^2}+\kappa \right)}{\omega_c}+\omega_c \omega_a} \end{pmatrix} \nonumber \\
\begin{pmatrix} \lambda_{x2} \\ \lambda_{y2} \end{pmatrix} & = \begin{pmatrix} \frac{1}{2} \sqrt{\frac{\sqrt{\kappa ^2 \omega_a^2 \left(\kappa ^2+\omega_c^2\right)}}{\omega_c}+\frac{\kappa ^2 \omega_a}{\omega_c}+\omega_c \omega_a} \\ \frac{1}{2} \sqrt{\frac{\omega_a \left(-\kappa  \sqrt{\kappa ^2+\omega_c^2}+\kappa ^2+\omega_c^2\right)}{\omega_c}} \end{pmatrix}\,. \nonumber
\end{align}
In order to distinguish which solutions correspond to a tricritical point we required $\lambda_x,\lambda_y$ to be positive real numbers for the parameters used in the main text. We can now evaluate the distance between these two points
\begin{align}
  \delta & = \sqrt{\left|\lambda_{x1}-\lambda_{x2}\right|^2 + \left|\lambda_{y1}-\lambda_{y2}\right|^2} \nonumber \\
        & = \sqrt{\omega_a \left(\frac{\kappa ^2}{\omega_c}+\omega_c-\sqrt{\kappa ^2+\omega_c^2}\right)}\,,
\end{align}
that for $\omega = \omega_c = \omega_a$ reduces to
\begin{equation}
  \label{resonancedelta}
  \delta = \omega \sqrt{1 + \frac{\kappa ^2}{\omega ^2}-\sqrt{1 + \frac{\kappa ^2}{\omega ^2}}}.
\end{equation}
\begin{figure}
    \includegraphics[width=\columnwidth]{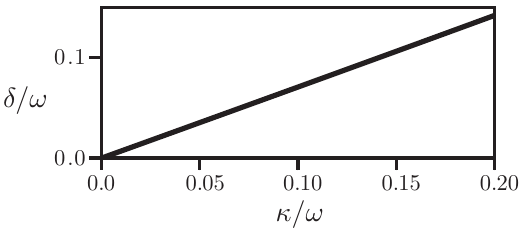}
    \caption{%
    Separation between the two out-of-equilibrium tricritical points, $\delta$, as a function of $\kappa$ for $\omega = \omega_c = \omega_a$, Eq.~\eqref{resonancedelta}. 
    }
    \label{fig1supmat}
\end{figure}
Expanding this expression for small $\kappa$ values $\kappa\ll\omega$ we get
\begin{equation}
  \delta = \frac{\kappa }{\sqrt{2} } + \mathcal{O}\left(\kappa ^2\right)\,.
\end{equation}

\section{V. Large-$N$ expansion} 
\setcounter{enumi}{5}
\setcounter{equation}{0}
Taking the IDTC model [Eq.~(1) in the main text] and applying the Holstein-Primakoff transformation for the spins, $S_+=  b^\dagger \sqrt{N - b^\dagger b}$ and $S_z= -\frac{N}{2} + b^\dagger b$, we obtain
\begin{align}
\label{eq:IDTC-HP}
H  &= \hbar\omega_c a^\dagger a + \hbar\omega_a  \left(-\frac{N}{2} + b^\dagger b\right) +\\
&+ {\frac{\hbar\lambda_x}{\sqrt{N}}} \left(b^\dagger \sqrt{N - b^\dagger b} + \sqrt{N - b^\dagger b}b\right) \left(a + a^\dagger\right)\nonumber\\
&+ {\frac{\hbar\lambda_y}{\sqrt{N}}} \left(b^\dagger \sqrt{N - b^\dagger b} - \sqrt{N - b^\dagger b}b\right) \left(a - a^\dagger\right)\,.\nonumber
\end{align}
We can rewrite the operators in terms of a mean-field classical part plus quantum fluctuations, $a=\alpha\sqrt{N}+c$ and $b=\beta\sqrt{N}+d$. Inserting these expressions into Eq.~\eqref{eq:IDTC-HP}, we can perform a large-$N$ limit expansion, i.e., collect only terms that do not vanish nor diverge in this limit. We obtain a fluctuation Hamiltonian of the form of Eq.~(7) in the main text, where the prefactors are
\begin{widetext}
\begin{align}
\label{Omega}\Omega_a &=\omega_a - \lambda_x \frac{\alphare\betare\left(1+ 3\left(1-|\beta|^2\right)\right) }{\left(1-|\beta|^2\right)^{3/2}}-\lambda_y \frac{\alphaim\betaim\left(1+ 3\left(1-|\beta|^2\right)\right) }{\left(1-|\beta|^2\right)^{3/2}}\\
&=\omega_a + \frac{ 2 (-5 + 6 Z) (X \lambda_x \alphare - Y \lambda_y \alphaim)}{(1 - 2 Z)^2}\,,\nonumber \\
\label{Gamma1}    \Gamma_1&=\lambda_x\frac{ \left(1-|\beta|^2- \beta^*\betare\right)}{\sqrt{1-|\beta|^2}}+\lambda_y\frac{ \left(1-|\beta|^2- i\beta^*\betaim\right)}{\sqrt{1-|\beta|^2}}\\
        &=\frac{\lambda_x (- 4 X (X + i Y) + (1 - 2 Z)^2)  + \lambda_y(- 4 Y (- i X + Y) + (1 - 2 Z)^2)}{ \sqrt{2 - 4 Z}(1 - 2 Z)}\,,\nonumber\\
\label{Gamma2}     \Gamma_2 &=\lambda_x\frac{ \left(1-|\beta|^2- \beta^*\betare\right)}{\sqrt{1-|\beta|^2}}-\lambda_y\frac{ \left(1-|\beta|^2- i\beta^*\betaim\right)}{\sqrt{1-|\beta|^2}}\\
        &=\frac{\lambda_x (- 4 X (X + i Y) + (1 - 2 Z)^2)  - \lambda_y(- 4 Y (- i X + Y) + (1 - 2 Z)^2)}{ \sqrt{2 - 4 Z}(1 - 2 Z)}\,,\nonumber\\
\label{Gamma3} \Gamma_3 &=  -\beta\left[\lambda_x\alphare\frac{ \left(2-2|\beta|^2+ \beta\betare\right)}{2\left(1-|\beta|^2\right)^{3/2}}+i\lambda_y\alphaim\frac{ \left(2-2|\beta|^2- i\beta\betaim\right)}{2\left(1-|\beta|^2\right)^{3/2}}\right]\\
& =-\frac{ 2 (X - i Y) ((- 2 X Y + i (2 Y^2 + (1 - 2 Z)^2)) \lambda_y \alphaim + (2 X (X - i Y) + (1 - 2 Z)^2) \lambda_x \alphare}{(1 - 2 Z)^3}\,,\nonumber
\end{align}
\end{widetext}
and for the second equality we have used the relation between the classical Holstein-Primakoff spin and the mean-field solutions, $\beta=(X-iY)/\sqrt{1/2 - Z}$.

\section{VI. Fluctuation equations of motion} 
\setcounter{enumi}{6}
\setcounter{equation}{0}
We are interested in the equal-time two-operator correlation functions. Taking, for example, the photon number fluctuation $ \langle c^\dag(t) c(t)\rangle$, its time-dependence can be obtained using Heisenberg's equation of motion
\begin{align}\label{heis}
\frac{d}{dt}  c^\dag(t) c(t) =&  -\frac{i}{\hbar} [H_{\rm fl}, c^\dag c]  - 2\kappa c^\dag c\,,
\end{align}
where we have also taken the large-$N$ limit of the Lindblad terms in the Liovillian [Eq.~(2) in the main text]. 

Note that the commutation of the photon number with $H_{\rm fl}$ couples it to the other two-operator correlation functions. Hence, in order to solve the time-evolution of the photon number, we need to evaluate the Heisenberg's equation of motion for all ten possible two-point correlations functions, and obtain
\begin{widetext}
\begin{align}
\frac{d}{dt} c^\dag c^\pdag =&- i \Gamma_1 c^\dagger d^\pdag + i \Gamma_1^* c^\pdag d^\dagger + i \Gamma_2 c^\pdag d^\pdag - i c^\dagger d^\dagger \Gamma_2^* - 2 \kappa c^\dagger c^\pdag \,,\\
\frac{d}{dt}  c^\pdag c^\pdag =& - 2 i \omega_c c^\pdag c^\pdag  -2 i \Gamma_1 c^\pdag d^\pdag - 2 i \Gamma_2^* c^\pdag d^\dag  - 2 \kappa c^\pdag c^\pdag \,,\\
\frac{d}{dt} c^\dag c^\dag =& 2 i \omega_c c^\dag c^\dag + 2 i \Gamma_1^* c^\dag d^\dag + 2 i \Gamma_2 c^\dag d^\pdag - 2 \kappa c^\dag c^\dag \,,\\
\frac{d}{dt} c^\pdag d^\pdag =& - i (\omega_c+\omega_a) c^\pdag d^\pdag - i \Gamma_1 d^\pdag d^\pdag - i \Gamma_1^* c^\pdag c^\pdag - i \Gamma_2^*\left(c^\dag c^\pdag + d^\dag d^\pdag+1\right) - 2 i \Gamma_3 c^\pdag d^\dag - \kappa c^\pdag d^\pdag\,,\\
\frac{d}{dt} c^\dag d^\dag =& i (\omega_c+\omega_a) c^\dag d^\dag + i \Gamma_1^* d^\dag d^\dag + i \Gamma_1 c^\dag c^\dag + i \Gamma_2 \left(c^\dag c^\pdag + d^\dag d^\pdag+1\right) + 2 i \Gamma_3^* c^\dag d^\pdag - \kappa c^\dag d^\dag \,,\\
\frac{d}{dt} c^\pdag d^\dag =&  i c^\pdag d^\dag (\omega_a - \omega_c ) + i \Gamma_1 \left(c^\dag c^\pdag - d^\dag d^\pdag\right) + i \Gamma_2 c^\pdag c^\pdag  - i \Gamma_2^* d^\dag d^\dag  +  2 i \Gamma_3^* c^\pdag d^\pdag  - \kappa c^\pdag d^\dag  \,,\\
\frac{d}{dt} c^\dag d^\pdag =& i c^\dag d^\pdag (\omega_c - \omega_a ) - i \Gamma_1^* \left(c^\dag c^\pdag - d^\dag d^\pdag\right) - i \Gamma_2^* c^\dag c^\dag  + i \Gamma_2 d^\pdag d^\pdag  -  2 i \Gamma_3 c^\dag d^\dag  - \kappa c^\dag d^\pdag \,,\\
\frac{d}{dt} d^\pdag d^\pdag =& - 2 i \omega_a d^\pdag d^\pdag - 2 i \Gamma_1^* c^\pdag d^\pdag - 2 i \Gamma_2^* c^\dag d^\pdag - 2 i \Gamma_3\left(1 + 2 d^\dag d^\pdag\right) \,,\\
\frac{d}{dt} d^\dag d^\dag =& 2 i \omega_a d^\dag d^\dag + 2 i \Gamma_1 c^\dag d^\dag + 2 i \Gamma_2 c^\pdag d^\dag + 2 i \Gamma_3^*\left(1 + 2 d^\dag d^\pdag\right)\,,\\
\frac{d}{dt} d^\dag d^\pdag =& i \Gamma_1 c^\dag d^\pdag - i\Gamma_1^* c^\pdag d^\dag + i \Gamma_2 c^\pdag d^\pdag - i \Gamma_2^* c^\dag d^\dag  - 2 i \Gamma_3 d^\dag d^\dag + 2 i \Gamma_3^* d^\pdag d^\pdag \,.
\end{align}
\end{widetext}
In the steady-state limit, the l.h.s. is set to zero and we obtain a set of ten coupled homogeneous equations. In the IDTC model, for all values $\lambda_x\neq 0$ and/or $\lambda_y\neq 0$, $\Gamma_1\neq 0$ [cf.~Eqs.~\eqref{x2}-\eqref{y2} and \eqref{Gamma1}]. The set of ten equations in invertible whenever $\Gamma_2\neq 0$ and/or $\Gamma_3\neq 0$, resulting in an analytical expression for all ten correlators in the steady-state limit. For brevity, we only write the photon number fluctuation
\begin{widetext}
\begin{align}
&\ave{c^\dag c}=\\
&\resizebox{1.\hsize}{!}{$\displaystyle \frac{2  \left( \kappa^2 + \omega_c^2 + \left|\Gamma_1\right|^2 - \left|\Gamma_2\right|^2 \right) 
               \left[ 2  \re\left\{\Gamma_1^2  \Gamma_2^2  \Gamma_3^2\right\} 
                 - 2  \omega_a  \re\left\{\Gamma_1  \Gamma_2  \Gamma_3\right\} 
                  \left(\left|\Gamma_1\right|^2 + \left|\Gamma_2\right|^2\right)
                 + \omega_a^2  \left|\Gamma_1\right|^2  \left|\Gamma_2\right|^2 
                 + \left|\Gamma_1\right|^4  \left|\Gamma_3\right|^2
                 + \left|\Gamma_2\right|^4  \left|\Gamma_3\right|^2 \right] }
              { \left[\left|\Gamma_1\right|^2  \left(\kappa^2 + (\omega_c + \omega_a)^2 - 4  \left|\Gamma_3\right|^2\right)
                  - \left|\Gamma_2\right|^2  \left(\kappa^2 + (\omega_c - \omega_a)^2 - 4  \left|\Gamma_3\right|^2\right) 
                  - 8  \omega_c  \re\left\{\Gamma_1  \Gamma_2  \Gamma_3\right\} \right]
                 \left[( \kappa^2 + \omega_c^2)  \left(\omega_a^2 - 4  \left|\Gamma_3\right|^2\right)
                   - 2  \omega_c  \omega_a  \left(\left|\Gamma_1\right|^2 + \left|\Gamma_2\right|^2\right)
                   + \left(\left|\Gamma_1\right|^2  - \left|\Gamma_2\right|^2\right)^2                   
                   + 8  \omega_c  \re\left\{\Gamma_1  \Gamma_2  \Gamma_3\right\} \right]  }\,,$}\nonumber
\end{align}
\end{widetext}
which in the normal phase simplifies to Eq.~(8) in the main text.


\begin{thebibliography}{42}%
\makeatletter
\providecommand \@ifxundefined [1]{%
 \@ifx{#1\undefined}
}%
\providecommand \@ifnum [1]{%
 \ifnum #1\expandafter \@firstoftwo
 \else \expandafter \@secondoftwo
 \fi
}%
\providecommand \@ifx [1]{%
 \ifx #1\expandafter \@firstoftwo
 \else \expandafter \@secondoftwo
 \fi
}%
\providecommand \natexlab [1]{#1}%
\providecommand \enquote  [1]{``#1''}%
\providecommand \bibnamefont  [1]{#1}%
\providecommand \bibfnamefont [1]{#1}%
\providecommand \citenamefont [1]{#1}%
\providecommand \href@noop [0]{\@secondoftwo}%
\providecommand \href [0]{\begingroup \@sanitize@url \@href}%
\providecommand \@href[1]{\@@startlink{#1}\@@href}%
\providecommand \@@href[1]{\endgroup#1\@@endlink}%
\providecommand \@sanitize@url [0]{\catcode `\\12\catcode `\$12\catcode
  `\&12\catcode `\#12\catcode `\^12\catcode `\_12\catcode `\%12\relax}%
\providecommand \@@startlink[1]{}%
\providecommand \@@endlink[0]{}%
\providecommand \url  [0]{\begingroup\@sanitize@url \@url }%
\providecommand \@url [1]{\endgroup\@href {#1}{\urlprefix }}%
\providecommand \urlprefix  [0]{URL }%
\providecommand \Eprint [0]{\href }%
\providecommand \doibase [0]{http://dx.doi.org/}%
\providecommand \selectlanguage [0]{\@gobble}%
\providecommand \bibinfo  [0]{\@secondoftwo}%
\providecommand \bibfield  [0]{\@secondoftwo}%
\providecommand \translation [1]{[#1]}%
\providecommand \BibitemOpen [0]{}%
\providecommand \bibitemStop [0]{}%
\providecommand \bibitemNoStop [0]{.\EOS\space}%
\providecommand \EOS [0]{\spacefactor3000\relax}%
\providecommand \BibitemShut  [1]{\csname bibitem#1\endcsname}%
\let\auto@bib@innerbib\@empty
\bibitem [{\citenamefont {Syassen}\ \emph {et~al.}(2008)\citenamefont
  {Syassen}, \citenamefont {Bauer}, \citenamefont {Lettner}, \citenamefont
  {Volz}, \citenamefont {Dietze}, \citenamefont {Garc{\'\i}a-Ripoll},
  \citenamefont {Cirac}, \citenamefont {Rempe},\ and\ \citenamefont
  {D{\"u}rr}}]{Syassen1329}%
  \BibitemOpen
  \bibfield  {author} {\bibinfo {author} {\bibfnamefont {N.}~\bibnamefont
  {Syassen}}, \bibinfo {author} {\bibfnamefont {D.~M.}\ \bibnamefont {Bauer}},
  \bibinfo {author} {\bibfnamefont {M.}~\bibnamefont {Lettner}}, \bibinfo
  {author} {\bibfnamefont {T.}~\bibnamefont {Volz}}, \bibinfo {author}
  {\bibfnamefont {D.}~\bibnamefont {Dietze}}, \bibinfo {author} {\bibfnamefont
  {J.~J.}\ \bibnamefont {Garc{\'\i}a-Ripoll}}, \bibinfo {author} {\bibfnamefont
  {J.~I.}\ \bibnamefont {Cirac}}, \bibinfo {author} {\bibfnamefont
  {G.}~\bibnamefont {Rempe}}, \ and\ \bibinfo {author} {\bibfnamefont
  {S.}~\bibnamefont {D{\"u}rr}},\ }\href {\doibase 10.1126/science.1155309}
  {\bibfield  {journal} {\bibinfo  {journal} {Science}\ }\textbf {\bibinfo
  {volume} {320}},\ \bibinfo {pages} {1329} (\bibinfo {year}
  {2008})}\BibitemShut {NoStop}%
\bibitem [{\citenamefont {Baumann}\ \emph {et~al.}(2010)\citenamefont
  {Baumann}, \citenamefont {Guerlin}, \citenamefont {Brennecke},\ and\
  \citenamefont {Esslinger}}]{Baumann2010}%
  \BibitemOpen
  \bibfield  {author} {\bibinfo {author} {\bibfnamefont {K.}~\bibnamefont
  {Baumann}}, \bibinfo {author} {\bibfnamefont {C.}~\bibnamefont {Guerlin}},
  \bibinfo {author} {\bibfnamefont {F.}~\bibnamefont {Brennecke}}, \ and\
  \bibinfo {author} {\bibfnamefont {T.}~\bibnamefont {Esslinger}},\ }\href
  {http://dx.doi.org/10.1038/nature09009} {\bibfield  {journal} {\bibinfo
  {journal} {Nature}\ }\textbf {\bibinfo {volume} {464}},\ \bibinfo {pages}
  {1301} (\bibinfo {year} {2010})}\BibitemShut {NoStop}%
\bibitem [{\citenamefont {Barreiro}\ \emph {et~al.}(2011)\citenamefont
  {Barreiro}, \citenamefont {M{\"u}ller}, \citenamefont {Schindler},
  \citenamefont {Nigg}, \citenamefont {Monz}, \citenamefont {Chwalla},
  \citenamefont {Hennrich}, \citenamefont {Roos}, \citenamefont {Zoller},\ and\
  \citenamefont {Blatt}}]{Barreiro2011}%
  \BibitemOpen
  \bibfield  {author} {\bibinfo {author} {\bibfnamefont {J.~T.}\ \bibnamefont
  {Barreiro}}, \bibinfo {author} {\bibfnamefont {M.}~\bibnamefont
  {M{\"u}ller}}, \bibinfo {author} {\bibfnamefont {P.}~\bibnamefont
  {Schindler}}, \bibinfo {author} {\bibfnamefont {D.}~\bibnamefont {Nigg}},
  \bibinfo {author} {\bibfnamefont {T.}~\bibnamefont {Monz}}, \bibinfo {author}
  {\bibfnamefont {M.}~\bibnamefont {Chwalla}}, \bibinfo {author} {\bibfnamefont
  {M.}~\bibnamefont {Hennrich}}, \bibinfo {author} {\bibfnamefont {C.~F.}\
  \bibnamefont {Roos}}, \bibinfo {author} {\bibfnamefont {P.}~\bibnamefont
  {Zoller}}, \ and\ \bibinfo {author} {\bibfnamefont {R.}~\bibnamefont
  {Blatt}},\ }\href {http://dx.doi.org/10.1038/nature09801} {\bibfield
  {journal} {\bibinfo  {journal} {Nature}\ }\textbf {\bibinfo {volume} {470}},\
  \bibinfo {pages} {486} (\bibinfo {year} {2011})}\BibitemShut {NoStop}%
\bibitem [{\citenamefont {Barontini}\ \emph {et~al.}(2013)\citenamefont
  {Barontini}, \citenamefont {Labouvie}, \citenamefont {Stubenrauch},
  \citenamefont {Vogler}, \citenamefont {Guarrera},\ and\ \citenamefont
  {Ott}}]{PhysRevLett.110.035302}%
  \BibitemOpen
  \bibfield  {author} {\bibinfo {author} {\bibfnamefont {G.}~\bibnamefont
  {Barontini}}, \bibinfo {author} {\bibfnamefont {R.}~\bibnamefont {Labouvie}},
  \bibinfo {author} {\bibfnamefont {F.}~\bibnamefont {Stubenrauch}}, \bibinfo
  {author} {\bibfnamefont {A.}~\bibnamefont {Vogler}}, \bibinfo {author}
  {\bibfnamefont {V.}~\bibnamefont {Guarrera}}, \ and\ \bibinfo {author}
  {\bibfnamefont {H.}~\bibnamefont {Ott}},\ }\href {\doibase
  10.1103/PhysRevLett.110.035302} {\bibfield  {journal} {\bibinfo  {journal}
  {Phys. Rev. Lett.}\ }\textbf {\bibinfo {volume} {110}},\ \bibinfo {pages}
  {035302} (\bibinfo {year} {2013})}\BibitemShut {NoStop}%
\bibitem [{\citenamefont {Carusotto}\ and\ \citenamefont
  {Ciuti}(2013)}]{Carusotto2013}%
  \BibitemOpen
  \bibfield  {author} {\bibinfo {author} {\bibfnamefont {I.}~\bibnamefont
  {Carusotto}}\ and\ \bibinfo {author} {\bibfnamefont {C.}~\bibnamefont
  {Ciuti}},\ }\href {\doibase 10.1103/RevModPhys.85.299} {\bibfield  {journal}
  {\bibinfo  {journal} {Rev. Mod. Phys.}\ }\textbf {\bibinfo {volume} {85}},\
  \bibinfo {pages} {299} (\bibinfo {year} {2013})}\BibitemShut {NoStop}%
\bibitem [{\citenamefont {Hartmann}\ \emph {et~al.}(2008)\citenamefont
  {Hartmann}, \citenamefont {Brand{\~a}o},\ and\ \citenamefont
  {Plenio}}]{Hartmann2008}%
  \BibitemOpen
  \bibfield  {author} {\bibinfo {author} {\bibfnamefont {M.}~\bibnamefont
  {Hartmann}}, \bibinfo {author} {\bibfnamefont {F.}~\bibnamefont
  {Brand{\~a}o}}, \ and\ \bibinfo {author} {\bibfnamefont {M.}~\bibnamefont
  {Plenio}},\ }\href {\doibase 10.1002/lpor.200810046} {\bibfield  {journal}
  {\bibinfo  {journal} {Laser {\&} Photonics Review}\ }\textbf {\bibinfo
  {volume} {2}},\ \bibinfo {pages} {527} (\bibinfo {year} {2008})}\BibitemShut
  {NoStop}%
\bibitem [{\citenamefont {Duan}\ \emph {et~al.}(2001)\citenamefont {Duan},
  \citenamefont {Lukin}, \citenamefont {Cirac},\ and\ \citenamefont
  {Zoller}}]{Zoller2001}%
  \BibitemOpen
  \bibfield  {author} {\bibinfo {author} {\bibfnamefont {L.~M.}\ \bibnamefont
  {Duan}}, \bibinfo {author} {\bibfnamefont {M.~D.}\ \bibnamefont {Lukin}},
  \bibinfo {author} {\bibfnamefont {J.~I.}\ \bibnamefont {Cirac}}, \ and\
  \bibinfo {author} {\bibfnamefont {P.}~\bibnamefont {Zoller}},\ }\href
  {http://dx.doi.org/10.1038/35106500} {\bibfield  {journal} {\bibinfo
  {journal} {Nature}\ }\textbf {\bibinfo {volume} {414}},\ \bibinfo {pages}
  {413} (\bibinfo {year} {2001})}\BibitemShut {NoStop}%
\bibitem [{\citenamefont {Bohnet}\ \emph {et~al.}(2012)\citenamefont {Bohnet},
  \citenamefont {Chen}, \citenamefont {Weiner}, \citenamefont {Meiser},
  \citenamefont {Holland},\ and\ \citenamefont {Thompson}}]{Bohnet2012}%
  \BibitemOpen
  \bibfield  {author} {\bibinfo {author} {\bibfnamefont {J.~G.}\ \bibnamefont
  {Bohnet}}, \bibinfo {author} {\bibfnamefont {Z.}~\bibnamefont {Chen}},
  \bibinfo {author} {\bibfnamefont {J.~M.}\ \bibnamefont {Weiner}}, \bibinfo
  {author} {\bibfnamefont {D.}~\bibnamefont {Meiser}}, \bibinfo {author}
  {\bibfnamefont {M.~J.}\ \bibnamefont {Holland}}, \ and\ \bibinfo {author}
  {\bibfnamefont {J.~K.}\ \bibnamefont {Thompson}},\ }\href
  {http://dx.doi.org/10.1038/nature10920} {\bibfield  {journal} {\bibinfo
  {journal} {Nature}\ }\textbf {\bibinfo {volume} {484}},\ \bibinfo {pages}
  {78} (\bibinfo {year} {2012})}\BibitemShut {NoStop}%
\bibitem [{\citenamefont {Fausti}\ \emph {et~al.}(2011)\citenamefont {Fausti},
  \citenamefont {Tobey}, \citenamefont {Dean}, \citenamefont {Kaiser},
  \citenamefont {Dienst}, \citenamefont {Hoffmann}, \citenamefont {Pyon},
  \citenamefont {Takayama}, \citenamefont {Takagi},\ and\ \citenamefont
  {Cavalleri}}]{cavalleri}%
  \BibitemOpen
  \bibfield  {author} {\bibinfo {author} {\bibfnamefont {D.}~\bibnamefont
  {Fausti}}, \bibinfo {author} {\bibfnamefont {R.~I.}\ \bibnamefont {Tobey}},
  \bibinfo {author} {\bibfnamefont {N.}~\bibnamefont {Dean}}, \bibinfo {author}
  {\bibfnamefont {S.}~\bibnamefont {Kaiser}}, \bibinfo {author} {\bibfnamefont
  {A.}~\bibnamefont {Dienst}}, \bibinfo {author} {\bibfnamefont {M.~C.}\
  \bibnamefont {Hoffmann}}, \bibinfo {author} {\bibfnamefont {S.}~\bibnamefont
  {Pyon}}, \bibinfo {author} {\bibfnamefont {T.}~\bibnamefont {Takayama}},
  \bibinfo {author} {\bibfnamefont {H.}~\bibnamefont {Takagi}}, \ and\ \bibinfo
  {author} {\bibfnamefont {A.}~\bibnamefont {Cavalleri}},\ }\href {\doibase
  10.1126/science.1197294} {\bibfield  {journal} {\bibinfo  {journal}
  {Science}\ }\textbf {\bibinfo {volume} {331}},\ \bibinfo {pages} {189}
  (\bibinfo {year} {2011})}\BibitemShut {NoStop}%
\bibitem [{\citenamefont {Tomadin}\ \emph {et~al.}(2011)\citenamefont
  {Tomadin}, \citenamefont {Diehl},\ and\ \citenamefont
  {Zoller}}]{Tomadin2011}%
  \BibitemOpen
  \bibfield  {author} {\bibinfo {author} {\bibfnamefont {A.}~\bibnamefont
  {Tomadin}}, \bibinfo {author} {\bibfnamefont {S.}~\bibnamefont {Diehl}}, \
  and\ \bibinfo {author} {\bibfnamefont {P.}~\bibnamefont {Zoller}},\ }\href
  {\doibase 10.1103/PhysRevA.83.013611} {\bibfield  {journal} {\bibinfo
  {journal} {Phys. Rev. A}\ }\textbf {\bibinfo {volume} {83}},\ \bibinfo
  {pages} {013611} (\bibinfo {year} {2011})}\BibitemShut {NoStop}%
\bibitem [{\citenamefont {Diehl}\ \emph {et~al.}(2010)\citenamefont {Diehl},
  \citenamefont {Tomadin}, \citenamefont {Micheli}, \citenamefont {Fazio},\
  and\ \citenamefont {Zoller}}]{Diehl2010}%
  \BibitemOpen
  \bibfield  {author} {\bibinfo {author} {\bibfnamefont {S.}~\bibnamefont
  {Diehl}}, \bibinfo {author} {\bibfnamefont {A.}~\bibnamefont {Tomadin}},
  \bibinfo {author} {\bibfnamefont {A.}~\bibnamefont {Micheli}}, \bibinfo
  {author} {\bibfnamefont {R.}~\bibnamefont {Fazio}}, \ and\ \bibinfo {author}
  {\bibfnamefont {P.}~\bibnamefont {Zoller}},\ }\href {\doibase
  10.1103/PhysRevLett.105.015702} {\bibfield  {journal} {\bibinfo  {journal}
  {Phys. Rev. Lett.}\ }\textbf {\bibinfo {volume} {105}},\ \bibinfo {pages}
  {015702} (\bibinfo {year} {2010})}\BibitemShut {NoStop}%
\bibitem [{\citenamefont {Nagy}\ \emph {et~al.}(2011)\citenamefont {Nagy},
  \citenamefont {Szirmai},\ and\ \citenamefont {Domokos}}]{Nagy2011}%
  \BibitemOpen
  \bibfield  {author} {\bibinfo {author} {\bibfnamefont {D.}~\bibnamefont
  {Nagy}}, \bibinfo {author} {\bibfnamefont {G.}~\bibnamefont {Szirmai}}, \
  and\ \bibinfo {author} {\bibfnamefont {P.}~\bibnamefont {Domokos}},\ }\href
  {\doibase 10.1103/PhysRevA.84.043637} {\bibfield  {journal} {\bibinfo
  {journal} {Phys. Rev. A}\ }\textbf {\bibinfo {volume} {84}},\ \bibinfo
  {pages} {043637} (\bibinfo {year} {2011})}\BibitemShut {NoStop}%
\bibitem [{\citenamefont {Diehl}\ \emph {et~al.}(2011)\citenamefont {Diehl},
  \citenamefont {Rico}, \citenamefont {Baranov},\ and\ \citenamefont
  {Zoller}}]{Diehl2011}%
  \BibitemOpen
  \bibfield  {author} {\bibinfo {author} {\bibfnamefont {S.}~\bibnamefont
  {Diehl}}, \bibinfo {author} {\bibfnamefont {E.}~\bibnamefont {Rico}},
  \bibinfo {author} {\bibfnamefont {M.~A.}\ \bibnamefont {Baranov}}, \ and\
  \bibinfo {author} {\bibfnamefont {P.}~\bibnamefont {Zoller}},\ }\href
  {http://dx.doi.org/10.1038/nphys2106} {\bibfield  {journal} {\bibinfo
  {journal} {Nature Physics}\ }\textbf {\bibinfo {volume} {7}},\ \bibinfo
  {pages} {971} (\bibinfo {year} {2011})}\BibitemShut {NoStop}%
\bibitem [{\citenamefont {Zhiqiang}\ \emph {et~al.}(2017)\citenamefont
  {Zhiqiang}, \citenamefont {Lee}, \citenamefont {Kumar}, \citenamefont
  {Arnold}, \citenamefont {Masson}, \citenamefont {Parkins},\ and\
  \citenamefont {Barrett}}]{Barrett2017}%
  \BibitemOpen
  \bibfield  {author} {\bibinfo {author} {\bibfnamefont {Z.}~\bibnamefont
  {Zhiqiang}}, \bibinfo {author} {\bibfnamefont {C.~H.}\ \bibnamefont {Lee}},
  \bibinfo {author} {\bibfnamefont {R.}~\bibnamefont {Kumar}}, \bibinfo
  {author} {\bibfnamefont {K.~J.}\ \bibnamefont {Arnold}}, \bibinfo {author}
  {\bibfnamefont {S.~J.}\ \bibnamefont {Masson}}, \bibinfo {author}
  {\bibfnamefont {A.~S.}\ \bibnamefont {Parkins}}, \ and\ \bibinfo {author}
  {\bibfnamefont {M.~D.}\ \bibnamefont {Barrett}},\ }\href {\doibase
  10.1364/OPTICA.4.000424} {\bibfield  {journal} {\bibinfo  {journal} {Optica}\
  }\textbf {\bibinfo {volume} {4}},\ \bibinfo {pages} {424} (\bibinfo {year}
  {2017})}\BibitemShut {NoStop}%
\bibitem [{\citenamefont {Baksic}\ and\ \citenamefont
  {Ciuti}(2014)}]{Baksic2014}%
  \BibitemOpen
  \bibfield  {author} {\bibinfo {author} {\bibfnamefont {A.}~\bibnamefont
  {Baksic}}\ and\ \bibinfo {author} {\bibfnamefont {C.}~\bibnamefont {Ciuti}},\
  }\href {\doibase 10.1103/PhysRevLett.112.173601} {\bibfield  {journal}
  {\bibinfo  {journal} {Phys. Rev. Lett.}\ }\textbf {\bibinfo {volume} {112}},\
  \bibinfo {pages} {173601} (\bibinfo {year} {2014})}\BibitemShut {NoStop}%
\bibitem [{sup()}]{supmat}%
  \BibitemOpen
  \href@noop {} {}\bibinfo {note} {See Supplemental Material for additional
  details, which includes Ref.~\cite{Bhaseen2012}}\BibitemShut {NoStop}%
\bibitem [{\citenamefont {Dimer}\ \emph {et~al.}(2007)\citenamefont {Dimer},
  \citenamefont {Estienne}, \citenamefont {Parkins},\ and\ \citenamefont
  {Carmichael}}]{Dimer}%
  \BibitemOpen
  \bibfield  {author} {\bibinfo {author} {\bibfnamefont {F.}~\bibnamefont
  {Dimer}}, \bibinfo {author} {\bibfnamefont {B.}~\bibnamefont {Estienne}},
  \bibinfo {author} {\bibfnamefont {A.~S.}\ \bibnamefont {Parkins}}, \ and\
  \bibinfo {author} {\bibfnamefont {H.~J.}\ \bibnamefont {Carmichael}},\ }\href
  {\doibase 10.1103/PhysRevA.75.013804} {\bibfield  {journal} {\bibinfo
  {journal} {Phys. Rev. A}\ }\textbf {\bibinfo {volume} {75}},\ \bibinfo
  {pages} {013804} (\bibinfo {year} {2007})}\BibitemShut {NoStop}%
\bibitem [{\citenamefont {Nagy}\ \emph {et~al.}(2010)\citenamefont {Nagy},
  \citenamefont {K\'onya}, \citenamefont {Szirmai},\ and\ \citenamefont
  {Domokos}}]{Domokos}%
  \BibitemOpen
  \bibfield  {author} {\bibinfo {author} {\bibfnamefont {D.}~\bibnamefont
  {Nagy}}, \bibinfo {author} {\bibfnamefont {G.}~\bibnamefont {K\'onya}},
  \bibinfo {author} {\bibfnamefont {G.}~\bibnamefont {Szirmai}}, \ and\
  \bibinfo {author} {\bibfnamefont {P.}~\bibnamefont {Domokos}},\ }\href
  {\doibase 10.1103/PhysRevLett.104.130401} {\bibfield  {journal} {\bibinfo
  {journal} {Phys. Rev. Lett.}\ }\textbf {\bibinfo {volume} {104}},\ \bibinfo
  {pages} {130401} (\bibinfo {year} {2010})}\BibitemShut {NoStop}%
\bibitem [{\citenamefont {Chirolli}\ \emph {et~al.}(2012)\citenamefont
  {Chirolli}, \citenamefont {Polini}, \citenamefont {Giovannetti},\ and\
  \citenamefont {MacDonald}}]{MacDonald}%
  \BibitemOpen
  \bibfield  {author} {\bibinfo {author} {\bibfnamefont {L.}~\bibnamefont
  {Chirolli}}, \bibinfo {author} {\bibfnamefont {M.}~\bibnamefont {Polini}},
  \bibinfo {author} {\bibfnamefont {V.}~\bibnamefont {Giovannetti}}, \ and\
  \bibinfo {author} {\bibfnamefont {A.~H.}\ \bibnamefont {MacDonald}},\ }\href
  {\doibase 10.1103/PhysRevLett.109.267404} {\bibfield  {journal} {\bibinfo
  {journal} {Phys. Rev. Lett.}\ }\textbf {\bibinfo {volume} {109}},\ \bibinfo
  {pages} {267404} (\bibinfo {year} {2012})}\BibitemShut {NoStop}%
\bibitem [{\citenamefont {Klinder}\ \emph {et~al.}(2015)\citenamefont
  {Klinder}, \citenamefont {Ke{\ss}ler}, \citenamefont {Wolke}, \citenamefont
  {Mathey},\ and\ \citenamefont {Hemmerich}}]{Klinder2015}%
  \BibitemOpen
  \bibfield  {author} {\bibinfo {author} {\bibfnamefont {J.}~\bibnamefont
  {Klinder}}, \bibinfo {author} {\bibfnamefont {H.}~\bibnamefont {Ke{\ss}ler}},
  \bibinfo {author} {\bibfnamefont {M.}~\bibnamefont {Wolke}}, \bibinfo
  {author} {\bibfnamefont {L.}~\bibnamefont {Mathey}}, \ and\ \bibinfo {author}
  {\bibfnamefont {A.}~\bibnamefont {Hemmerich}},\ }\href {\doibase
  10.1073/pnas.1417132112} {\bibfield  {journal} {\bibinfo  {journal}
  {Proceedings of the National Academy of Sciences}\ }\textbf {\bibinfo
  {volume} {112}},\ \bibinfo {pages} {3290} (\bibinfo {year}
  {2015})}\BibitemShut {NoStop}%
\bibitem [{\citenamefont {L{\'e}onard}\ \emph {et~al.}(2017)\citenamefont
  {L{\'e}onard}, \citenamefont {Morales}, \citenamefont {Zupancic},
  \citenamefont {Esslinger},\ and\ \citenamefont {Donner}}]{Leonard2017}%
  \BibitemOpen
  \bibfield  {author} {\bibinfo {author} {\bibfnamefont {J.}~\bibnamefont
  {L{\'e}onard}}, \bibinfo {author} {\bibfnamefont {A.}~\bibnamefont
  {Morales}}, \bibinfo {author} {\bibfnamefont {P.}~\bibnamefont {Zupancic}},
  \bibinfo {author} {\bibfnamefont {T.}~\bibnamefont {Esslinger}}, \ and\
  \bibinfo {author} {\bibfnamefont {T.}~\bibnamefont {Donner}},\ }\href
  {http://dx.doi.org/10.1038/nature21067} {\bibfield  {journal} {\bibinfo
  {journal} {Nature}\ }\textbf {\bibinfo {volume} {543}},\ \bibinfo {pages}
  {87} (\bibinfo {year} {2017})}\BibitemShut {NoStop}%
\bibitem [{\citenamefont {Morrison}\ and\ \citenamefont
  {Parkins}(2008)}]{Morrison2008}%
  \BibitemOpen
  \bibfield  {author} {\bibinfo {author} {\bibfnamefont {S.}~\bibnamefont
  {Morrison}}\ and\ \bibinfo {author} {\bibfnamefont {A.~S.}\ \bibnamefont
  {Parkins}},\ }\href {\doibase 10.1103/PhysRevLett.100.040403} {\bibfield
  {journal} {\bibinfo  {journal} {Phys. Rev. Lett.}\ }\textbf {\bibinfo
  {volume} {100}},\ \bibinfo {pages} {040403} (\bibinfo {year}
  {2008})}\BibitemShut {NoStop}%
\bibitem [{\citenamefont {Fan}\ \emph {et~al.}(2014)\citenamefont {Fan},
  \citenamefont {Yang}, \citenamefont {Zhang}, \citenamefont {Ma},
  \citenamefont {Chen},\ and\ \citenamefont {Jia}}]{Fan2014}%
  \BibitemOpen
  \bibfield  {author} {\bibinfo {author} {\bibfnamefont {J.}~\bibnamefont
  {Fan}}, \bibinfo {author} {\bibfnamefont {Z.}~\bibnamefont {Yang}}, \bibinfo
  {author} {\bibfnamefont {Y.}~\bibnamefont {Zhang}}, \bibinfo {author}
  {\bibfnamefont {J.}~\bibnamefont {Ma}}, \bibinfo {author} {\bibfnamefont
  {G.}~\bibnamefont {Chen}}, \ and\ \bibinfo {author} {\bibfnamefont
  {S.}~\bibnamefont {Jia}},\ }\href {\doibase 10.1103/PhysRevA.89.023812}
  {\bibfield  {journal} {\bibinfo  {journal} {Phys. Rev. A}\ }\textbf {\bibinfo
  {volume} {89}},\ \bibinfo {pages} {023812} (\bibinfo {year}
  {2014})}\BibitemShut {NoStop}%
\bibitem [{\citenamefont {Kirton}\ and\ \citenamefont
  {Keeling}(2017)}]{Keeling2017}%
  \BibitemOpen
  \bibfield  {author} {\bibinfo {author} {\bibfnamefont {P.}~\bibnamefont
  {Kirton}}\ and\ \bibinfo {author} {\bibfnamefont {J.}~\bibnamefont
  {Keeling}},\ }\href {\doibase 10.1103/PhysRevLett.118.123602} {\bibfield
  {journal} {\bibinfo  {journal} {Phys. Rev. Lett.}\ }\textbf {\bibinfo
  {volume} {118}},\ \bibinfo {pages} {123602} (\bibinfo {year}
  {2017})}\BibitemShut {NoStop}%
\bibitem [{\citenamefont {Larson}\ and\ \citenamefont
  {Irish}(2017)}]{Larson2017}%
  \BibitemOpen
  \bibfield  {author} {\bibinfo {author} {\bibfnamefont {J.}~\bibnamefont
  {Larson}}\ and\ \bibinfo {author} {\bibfnamefont {E.~K.}\ \bibnamefont
  {Irish}},\ }\href {http://stacks.iop.org/1751-8121/50/i=17/a=174002}
  {\bibfield  {journal} {\bibinfo  {journal} {Journal of Physics A:
  Mathematical and Theoretical}\ }\textbf {\bibinfo {volume} {50}},\ \bibinfo
  {pages} {174002} (\bibinfo {year} {2017})}\BibitemShut {NoStop}%
\bibitem [{\citenamefont {Nagy}\ and\ \citenamefont
  {Domokos}(2015)}]{Nagy2015}%
  \BibitemOpen
  \bibfield  {author} {\bibinfo {author} {\bibfnamefont {D.}~\bibnamefont
  {Nagy}}\ and\ \bibinfo {author} {\bibfnamefont {P.}~\bibnamefont {Domokos}},\
  }\href {\doibase 10.1103/PhysRevLett.115.043601} {\bibfield  {journal}
  {\bibinfo  {journal} {Phys. Rev. Lett.}\ }\textbf {\bibinfo {volume} {115}},\
  \bibinfo {pages} {043601} (\bibinfo {year} {2015})}\BibitemShut {NoStop}%
\bibitem [{\citenamefont {Kulkarni}\ \emph {et~al.}(2013)\citenamefont
  {Kulkarni}, \citenamefont {\"Oztop},\ and\ \citenamefont
  {T\"ureci}}]{Kulkarni2013}%
  \BibitemOpen
  \bibfield  {author} {\bibinfo {author} {\bibfnamefont {M.}~\bibnamefont
  {Kulkarni}}, \bibinfo {author} {\bibfnamefont {B.}~\bibnamefont {\"Oztop}}, \
  and\ \bibinfo {author} {\bibfnamefont {H.~E.}\ \bibnamefont {T\"ureci}},\
  }\href {\doibase 10.1103/PhysRevLett.111.220408} {\bibfield  {journal}
  {\bibinfo  {journal} {Phys. Rev. Lett.}\ }\textbf {\bibinfo {volume} {111}},\
  \bibinfo {pages} {220408} (\bibinfo {year} {2013})}\BibitemShut {NoStop}%
\bibitem [{\citenamefont {Brennecke}\ \emph {et~al.}(2013)\citenamefont
  {Brennecke}, \citenamefont {Mottl}, \citenamefont {Baumann}, \citenamefont
  {Landig}, \citenamefont {Donner},\ and\ \citenamefont
  {Esslinger}}]{Brennecke2013}%
  \BibitemOpen
  \bibfield  {author} {\bibinfo {author} {\bibfnamefont {F.}~\bibnamefont
  {Brennecke}}, \bibinfo {author} {\bibfnamefont {R.}~\bibnamefont {Mottl}},
  \bibinfo {author} {\bibfnamefont {K.}~\bibnamefont {Baumann}}, \bibinfo
  {author} {\bibfnamefont {R.}~\bibnamefont {Landig}}, \bibinfo {author}
  {\bibfnamefont {T.}~\bibnamefont {Donner}}, \ and\ \bibinfo {author}
  {\bibfnamefont {T.}~\bibnamefont {Esslinger}},\ }\href {\doibase
  10.1073/pnas.1306993110} {\bibfield  {journal} {\bibinfo  {journal}
  {Proceedings of the National Academy of Sciences}\ }\textbf {\bibinfo
  {volume} {110}},\ \bibinfo {pages} {11763} (\bibinfo {year}
  {2013})}\BibitemShut {NoStop}%
\bibitem [{\citenamefont {Keeling}\ \emph {et~al.}(2010)\citenamefont
  {Keeling}, \citenamefont {Bhaseen},\ and\ \citenamefont
  {Simons}}]{Keeling2010}%
  \BibitemOpen
  \bibfield  {author} {\bibinfo {author} {\bibfnamefont {J.}~\bibnamefont
  {Keeling}}, \bibinfo {author} {\bibfnamefont {M.~J.}\ \bibnamefont
  {Bhaseen}}, \ and\ \bibinfo {author} {\bibfnamefont {B.~D.}\ \bibnamefont
  {Simons}},\ }\href {\doibase 10.1103/PhysRevLett.105.043001} {\bibfield
  {journal} {\bibinfo  {journal} {Phys. Rev. Lett.}\ }\textbf {\bibinfo
  {volume} {105}},\ \bibinfo {pages} {043001} (\bibinfo {year}
  {2010})}\BibitemShut {NoStop}%
\bibitem [{\citenamefont {Griffiths}(1970)}]{griffiths}%
  \BibitemOpen
  \bibfield  {author} {\bibinfo {author} {\bibfnamefont {R.~B.}\ \bibnamefont
  {Griffiths}},\ }\href {https://link.aps.org/doi/10.1103/PhysRevLett.24.715}
  {\bibfield  {journal} {\bibinfo  {journal} {Physical Review Letters}\
  }\textbf {\bibinfo {volume} {24}},\ \bibinfo {pages} {715} (\bibinfo {year}
  {1970})}\BibitemShut {NoStop}%
\bibitem [{\citenamefont {Dicke}(1954)}]{Dicke}%
  \BibitemOpen
  \bibfield  {author} {\bibinfo {author} {\bibfnamefont {R.~H.}\ \bibnamefont
  {Dicke}},\ }\href {\doibase 10.1103/PhysRev.93.99} {\bibfield  {journal}
  {\bibinfo  {journal} {Phys. Rev.}\ }\textbf {\bibinfo {volume} {93}},\
  \bibinfo {pages} {99} (\bibinfo {year} {1954})}\BibitemShut {NoStop}%
\bibitem [{\citenamefont {Emary}\ and\ \citenamefont
  {Brandes}(2003)}]{Emary2003}%
  \BibitemOpen
  \bibfield  {author} {\bibinfo {author} {\bibfnamefont {C.}~\bibnamefont
  {Emary}}\ and\ \bibinfo {author} {\bibfnamefont {T.}~\bibnamefont
  {Brandes}},\ }\href {\doibase 10.1103/PhysRevE.67.066203} {\bibfield
  {journal} {\bibinfo  {journal} {Phys. Rev. E}\ }\textbf {\bibinfo {volume}
  {67}},\ \bibinfo {pages} {066203} (\bibinfo {year} {2003})}\BibitemShut
  {NoStop}%
\bibitem [{\citenamefont {Tavis}\ and\ \citenamefont
  {Cummings}(1968)}]{TavisCummings}%
  \BibitemOpen
  \bibfield  {author} {\bibinfo {author} {\bibfnamefont {M.}~\bibnamefont
  {Tavis}}\ and\ \bibinfo {author} {\bibfnamefont {F.~W.}\ \bibnamefont
  {Cummings}},\ }\href {\doibase 10.1103/PhysRev.170.379} {\bibfield  {journal}
  {\bibinfo  {journal} {Phys. Rev.}\ }\textbf {\bibinfo {volume} {170}},\
  \bibinfo {pages} {379} (\bibinfo {year} {1968})}\BibitemShut {NoStop}%
\bibitem [{\citenamefont {Vogel}\ and\ \citenamefont
  {Welsch}(2006)}]{vogel2006}%
  \BibitemOpen
  \bibfield  {author} {\bibinfo {author} {\bibfnamefont {W.}~\bibnamefont
  {Vogel}}\ and\ \bibinfo {author} {\bibfnamefont {D.-G.}\ \bibnamefont
  {Welsch}},\ }\href@noop {} {\emph {\bibinfo {title} {Quantum optics}}}\
  (\bibinfo  {publisher} {John Wiley {\&} Sons},\ \bibinfo {year}
  {2006})\BibitemShut {NoStop}%
\bibitem [{\citenamefont {Shirai}\ \emph {et~al.}(2014)\citenamefont {Shirai},
  \citenamefont {Mori},\ and\ \citenamefont {Miyashita}}]{Miyashita2014}%
  \BibitemOpen
  \bibfield  {author} {\bibinfo {author} {\bibfnamefont {T.}~\bibnamefont
  {Shirai}}, \bibinfo {author} {\bibfnamefont {T.}~\bibnamefont {Mori}}, \ and\
  \bibinfo {author} {\bibfnamefont {S.}~\bibnamefont {Miyashita}},\ }\href
  {http://stacks.iop.org/0953-4075/47/i=2/a=025501} {\bibfield  {journal}
  {\bibinfo  {journal} {Journal of Physics B: Atomic, Molecular and Optical
  Physics}\ }\textbf {\bibinfo {volume} {47}},\ \bibinfo {pages} {025501}
  (\bibinfo {year} {2014})}\BibitemShut {NoStop}%
\bibitem [{\citenamefont {Lolli}\ \emph {et~al.}(2015)\citenamefont {Lolli},
  \citenamefont {Baksic}, \citenamefont {Nagy}, \citenamefont {Manucharyan},\
  and\ \citenamefont {Ciuti}}]{Lolli2015}%
  \BibitemOpen
  \bibfield  {author} {\bibinfo {author} {\bibfnamefont {J.}~\bibnamefont
  {Lolli}}, \bibinfo {author} {\bibfnamefont {A.}~\bibnamefont {Baksic}},
  \bibinfo {author} {\bibfnamefont {D.}~\bibnamefont {Nagy}}, \bibinfo {author}
  {\bibfnamefont {V.~E.}\ \bibnamefont {Manucharyan}}, \ and\ \bibinfo {author}
  {\bibfnamefont {C.}~\bibnamefont {Ciuti}},\ }\href {\doibase
  10.1103/PhysRevLett.114.183601} {\bibfield  {journal} {\bibinfo  {journal}
  {Phys. Rev. Lett.}\ }\textbf {\bibinfo {volume} {114}},\ \bibinfo {pages}
  {183601} (\bibinfo {year} {2015})}\BibitemShut {NoStop}%
\bibitem [{dre()}]{dressed}%
  \BibitemOpen
  \href@noop {} {}\bibinfo {note} {The master equation description
  [Eq.~\eqref{liouvillian}] is valid in the Born-Markov limit of weak
  dissipation. In the limit of strong cavity-atom coupling it is advisable to
  use dressed operators for the Lindblad dissipators~\cite{Miyashita2014},
  which introduce new dissipation channels, e.g., spin decay and multi-photon
  processes. Nevertheless, the sensitivity of the Tavis-Cummings line to
  infinitesimal cavity-dissipation hints to the fact that dissipation in the
  ultra-strong-coupling regime would only affect our reported phenomena in a
  quantitative way.}\BibitemShut {Stop}%
\bibitem [{\citenamefont {Dalla~Torre}\ \emph {et~al.}(2016)\citenamefont
  {Dalla~Torre}, \citenamefont {Shchadilova}, \citenamefont {Wilner},
  \citenamefont {Lukin},\ and\ \citenamefont {Demler}}]{Torre2016}%
  \BibitemOpen
  \bibfield  {author} {\bibinfo {author} {\bibfnamefont {E.~G.}\ \bibnamefont
  {Dalla~Torre}}, \bibinfo {author} {\bibfnamefont {Y.}~\bibnamefont
  {Shchadilova}}, \bibinfo {author} {\bibfnamefont {E.~Y.}\ \bibnamefont
  {Wilner}}, \bibinfo {author} {\bibfnamefont {M.~D.}\ \bibnamefont {Lukin}}, \
  and\ \bibinfo {author} {\bibfnamefont {E.}~\bibnamefont {Demler}},\ }\href
  {\doibase 10.1103/PhysRevA.94.061802} {\bibfield  {journal} {\bibinfo
  {journal} {Phys. Rev. A}\ }\textbf {\bibinfo {volume} {94}},\ \bibinfo
  {pages} {061802} (\bibinfo {year} {2016})}\BibitemShut {NoStop}%
\bibitem [{\citenamefont {{\"O}ztop}\ \emph {et~al.}(2012)\citenamefont
  {{\"O}ztop}, \citenamefont {Bordyuh}, \citenamefont {M{\"u}stecaplıo{\u
  g}lu},\ and\ \citenamefont {T{\"u}reci}}]{Oztop2012}%
  \BibitemOpen
  \bibfield  {author} {\bibinfo {author} {\bibfnamefont {B.}~\bibnamefont
  {{\"O}ztop}}, \bibinfo {author} {\bibfnamefont {M.}~\bibnamefont {Bordyuh}},
  \bibinfo {author} {\bibfnamefont {{\"O}.~E.}\ \bibnamefont
  {M{\"u}stecaplıo{\u g}lu}}, \ and\ \bibinfo {author} {\bibfnamefont {H.~E.}\
  \bibnamefont {T{\"u}reci}},\ }\href
  {http://stacks.iop.org/1367-2630/14/i=8/a=085011} {\bibfield  {journal}
  {\bibinfo  {journal} {New Journal of Physics}\ }\textbf {\bibinfo {volume}
  {14}},\ \bibinfo {pages} {085011} (\bibinfo {year} {2012})}\BibitemShut
  {NoStop}%
\bibitem [{\citenamefont {Konczykowski}\ \emph {et~al.}(2006)\citenamefont
  {Konczykowski}, \citenamefont {van~der Beek}, \citenamefont {Koshelev},
  \citenamefont {Mosser}, \citenamefont {Dodgson},\ and\ \citenamefont
  {Kes}}]{vortex}%
  \BibitemOpen
  \bibfield  {author} {\bibinfo {author} {\bibfnamefont {M.}~\bibnamefont
  {Konczykowski}}, \bibinfo {author} {\bibfnamefont {C.~J.}\ \bibnamefont
  {van~der Beek}}, \bibinfo {author} {\bibfnamefont {A.~E.}\ \bibnamefont
  {Koshelev}}, \bibinfo {author} {\bibfnamefont {V.}~\bibnamefont {Mosser}},
  \bibinfo {author} {\bibfnamefont {M.}~\bibnamefont {Dodgson}}, \ and\
  \bibinfo {author} {\bibfnamefont {P.~H.}\ \bibnamefont {Kes}},\ }\href
  {\doibase 10.1103/PhysRevLett.97.237005} {\bibfield  {journal} {\bibinfo
  {journal} {Phys. Rev. Lett.}\ }\textbf {\bibinfo {volume} {97}},\ \bibinfo
  {pages} {237005} (\bibinfo {year} {2006})}\BibitemShut {NoStop}%
\bibitem [{\citenamefont {Moodie}\ \emph {et~al.}(2018)\citenamefont {Moodie},
  \citenamefont {Ballantine},\ and\ \citenamefont {Keeling}}]{Keeling2018}%
  \BibitemOpen
  \bibfield  {author} {\bibinfo {author} {\bibfnamefont {R.~I.}\ \bibnamefont
  {Moodie}}, \bibinfo {author} {\bibfnamefont {K.~E.}\ \bibnamefont
  {Ballantine}}, \ and\ \bibinfo {author} {\bibfnamefont {J.}~\bibnamefont
  {Keeling}},\ }\href {\doibase 10.1103/PhysRevA.97.033802} {\bibfield
  {journal} {\bibinfo  {journal} {Phys. Rev. A}\ }\textbf {\bibinfo {volume}
  {97}},\ \bibinfo {pages} {033802} (\bibinfo {year} {2018})}\BibitemShut
  {NoStop}%
\bibitem [{\citenamefont {Bhaseen}\ \emph {et~al.}(2012)\citenamefont
  {Bhaseen}, \citenamefont {Mayoh}, \citenamefont {Simons},\ and\ \citenamefont
  {Keeling}}]{Bhaseen2012}%
  \BibitemOpen
  \bibfield  {author} {\bibinfo {author} {\bibfnamefont {M.~J.}\ \bibnamefont
  {Bhaseen}}, \bibinfo {author} {\bibfnamefont {J.}~\bibnamefont {Mayoh}},
  \bibinfo {author} {\bibfnamefont {B.~D.}\ \bibnamefont {Simons}}, \ and\
  \bibinfo {author} {\bibfnamefont {J.}~\bibnamefont {Keeling}},\ }\href
  {\doibase 10.1103/PhysRevA.85.013817} {\bibfield  {journal} {\bibinfo
  {journal} {Phys. Rev. A}\ }\textbf {\bibinfo {volume} {85}},\ \bibinfo
  {pages} {013817} (\bibinfo {year} {2012})}\BibitemShut {NoStop}%
\end{thebibliography}
\end{document}